\documentclass{aastex701}

\usepackage{amsmath}
\usepackage{wasysym}
\usepackage{booktabs}
\usepackage{multirow}

\submitjournal{ApJs}

\begin{document}

\title{Multi-year forecasting of F10.7 and F30 solar radio flux using an adaptive Kalman filter and hemispheric solar activity}

\author[orcid=0009-0009-1264-4202,gname=Zaina, sname='Abu-Shaar']{Zaina Abu-Shaar}
\affiliation{Center for Engineering Systems and Sciences, Bolshoy Boulevard 30, bld. 1, Moscow 121205, Russia}
\email[show]{zaina.abushaar@ingsci.tech}  

\author[orcid=0000-0002-9189-1579,gname=Tatiana, sname='Podladchikova']{Tatiana Podladchikova} 
\affiliation{Center for Engineering Systems and Sciences, Bolshoy Boulevard 30, bld. 1, Moscow 121205, Russia}
\email[show]{tatiana.podladchikova@ingsci.tech}  

\author[orcid=0000-0003-2073-002X,gname=Astrid M., sname='Veronig']{Astrid M. Veronig}
\affiliation{Institute of Physics, University of Graz, Universit{\"a}tsplatz 5, A-8010 Graz, Austria}
\affiliation{Kanzelh{\"o}he Observatory for Solar and Environmental Research, University of Graz, Kanzelh{\"o}he 19, 9521, Treffen, Austria}
\email{astrid.veronig@uni-graz.at}

\author[orcid=0000-0001-7662-1960,gname=Stefan J., sname='Hofmeister']{Stefan J. Hofmeister}
\affiliation{Institute for Physics, ETH Zürich, Rämistrasse 101, 8092 Zürich, Switzerland}
\email{shofmeister@ethz.ch}  

\author[orcid=0000-0002-2117-9842,gname=Anatoly ., sname='Petrukovich']{Anatoly Petrukovich}
\affiliation{Independent Researcher}
\email{}

\begin{abstract}
The F10.7 and F30 radio flux indices drive thermospheric and radiation-environment models, and long-term predictions are needed for satellite mission planning, orbital-lifetime estimates, and re-entry assessments. We develop multi-year predictions of both indices for lead times of 1 to 84 months using three complementary contributions. First, we apply the McNish--Lincoln (M\&L) method and demonstrate its predictive skill over multi-year horizons, with RMSE of 11--22~sfu for F10.7 and 7--14~sfu for F30. RMSE largely stabilizes after roughly 36 months, indicating near-constant forecast skill over the seven-year horizon. Accuracy also depends strongly on the solar-cycle phase at which the forecast is issued, with the most favorable starting intervals occurring progressively earlier in the cycle for longer lead times. Second, we update the initial predictions using an adaptive Kalman filter with noise statistics identification, which assimilates the latest monthly observations to reconstruct the current smoothed flux, remove the delay introduced by centered smoothing, and improve solar-cycle state estimation, reducing forecast errors by up to 70\%. Third, we introduce synthetic hemispheric components of solar radio flux to account for the separate evolution of the northern and southern hemispheres, and incorporate them into a hemispheric M\&L formulation. Compared with the initial M\&L predictions, it reduces RMSE by up to 55\% for F10.7 and 58\% for F30, showing that hemispheric asymmetry contains predictive information lost in the full-disk index. These results extend solar-cycle forecasting to multi-year timescales relevant for long-term space-weather applications and operational planning.
\end{abstract}


\section{Introduction}
The evolution of solar activity on multi-year timescales determines the background environment in which satellites operate and space missions are planned. Solar ultraviolet and extreme-ultraviolet radiation change the temperature and density of the upper atmosphere and, consequently, the atmospheric drag on objects in low Earth orbit. Predictions of solar activity therefore enter the calculation of orbital lifetimes, re-entry epochs, and the long-term evolution of the debris population \citep{Virgili2014,Dolado2015,Virgili2017,Virgili2017ISO, Merz2017}. Such long-term predictions may become particularly valuable for the planning of satellite missions with lifetimes of several years and for the long-term operation of spacecraft already in orbit by improving estimates of future thermospheric conditions. Solar radio flux provides a long-term ground-based  measure of this evolution. In particular, the F10.7 and F30 indices serve as proxies for the solar radiative input to thermospheric models \citep{Tapping2013,Dudok2014}. Their predictive accuracy has practical consequences, as demonstrated by \citet{Petrova2021}, who showed that improved F10.7 predictions can also improve the predicted re-entry epochs in historical ESA re-entry campaigns.

The solar cycle also modulates the radiation environment in near-Earth space and in the atmosphere. Galactic cosmic rays (GCRs) are modulated during their transport through the heliosphere, and their intensity depends on the level and history of solar activity. Consequently, reliable long-term solar activity predictions are valuable for estimating the radiation environment over multi-year timescales \citep{Potgieter2013}.  Models of GCR spectra provide the incident particle environment for calculations of radiation exposure in space and in the atmosphere \citep{Matthia2014, Meier2018}. At aviation altitudes, the resulting secondary-particle field is used to estimate radiation exposure along flight routes \citep{Beck2007measurements, Latocha2009avidos}.

Different approaches have been developed for solar-cycle prediction, including precursor methods, physics-based models, and statistical techniques. Precursor methods estimate the amplitude of a future cycle from quantities measured earlier in the solar cycle evolution, including geomagnetic activity \citep{OhlOhl1979,Feynman1982}, polar magnetic fields \citep{SchattenSofia1987,Svalgaard2005,MunozJaramillo2012}, and solar activity indices \citep{CameronSchlussler2008,PodladchikovaLefebvreLinden2008,
Podladchikova2011,Podladchikova2012,Podladchikova2017}. Physics-based approaches, including dynamo and flux-transport models, describe the magnetic processes underlying the solar cycle \citep{Nandy2002,Dikpati2006,CameronSchlussler2015}. Statistical methods use the observed relationships between successive activity levels and between past cycles \citep{Macpherson1995,Conway1998,Sello2001,Kakad2020}. Data-driven approaches, including machine-learning and deep-learning techniques, have also been explored for solar-cycle and solar radio flux prediction \citep{Yaya2017,Benson2020,Stevenson2022,Jerse2022,Egorov2025}. These approaches and their limitations have been reviewed by \citet{Hathaway2010,Petrovay2020}.

For continuous long-term solar radio flux predictions, the goal is to estimate the evolution of the activity level month by month, including the rising phase, solar maximum, and declining phase of the cycle. The relevant predictive information can therefore depend both on the issue time within the cycle and on the forecast lead time. The McNish--Lincoln method \citep[hereafter M\&L;][]{Mcnish1949} uses the mean solar cycle to predict future activity. It adjusts the mean-cycle value at the forecast month based on the difference between the latest available activity level and the mean-cycle value at the corresponding month. This adjustment is determined by regression coefficients estimated from past solar cycles. Its application to solar radio flux was developed by \citet{Holland1984}. The difficulty arises because centered smoothing delays the last available smoothed observation relative to the current month. \citet{Podladchikova2012} addressed this difficulty for sunspot numbers with an adaptive Kalman filter. The filter assimilates recent monthly observations and estimates the smoothed activity up to the current time, which becomes the starting point of the updated prediction. This approach was implemented for operational forecasting and is used for monthly updated predictions of solar activity for the next 1--12 months at the World Data Center SILSO (\url{http:/sidc.be/silso}). \cite{Petrova2021} developed this approach for F10.7 and F30 in the RESONANCE method for 1--24 month-ahead forecasts. However, operational solar radio flux predictions have so far been limited mainly to short- and medium-term horizons, leaving multi-year prediction capabilities less explored. The recent analysis by \citet{Clette2024} showed that M\&L forecasting is sensitive to the cycle phase, with longer useful prediction horizons from the rising phase through the early declining phase and limitations near cycle boundaries.

A further source of predictive information is the separate evolution of the northern and southern hemispheres. \citet{Veronig2021} developed an extended hemispheric sunspot-number catalogue for 1874--2020 and demonstrated that hemispheric activity indicators provide additional information for solar-cycle prediction. \citet{Podladchikova2022} further showed that the maximal growth rates of the hemispheric sunspot series improve the prediction of the cycle amplitude compared with the corresponding full-disk indicator, demonstrating the predictive value of hemispheric information that captures the behaviour of solar activity and the solar magnetic field that may not evolve in phase in the Northern and Southern hemispheres.

In this study, we develop a multi-year forecasting framework for solar radio flux and derive predictions of the F10.7 and F30 indices for lead times from 1 to 84 months ahead. Building upon the RESONANCE approach of \citet{Petrova2021}, 
previously evaluated for forecasts up to 24 months ahead, we extend its capabilities to multi-year horizons by incorporating the McNish--Lincoln method with adaptive state estimation and hemispheric solar-cycle information. The preparation and reconstruction of the monthly mean radio-flux records are described in Section~\ref{data_preparation}. We first apply the M\&L method to obtain initial multi-year predictions (Section~\ref{initial_ML}). We then update these predictions using an adaptive Kalman filter, which reconstructs the current smoothed activity level and provides an improved basis for forecasting (Section~\ref{KFML}). The adaptive filter incorporates a new noise statistics identification procedure developed for solar radio flux forecasting, which is essential for reliable state estimation (Section~\ref{KFML}; Appendix~\ref{appendix_NSI}). Finally, we assess the predictive value of hemispheric sunspot evolution for multi-year forecast accuracy by comparing predictions based on total solar activity with those derived from separate hemispheric components (Section~\ref{Hemispheric}). The results are discussed and summarized in Section~\ref{Discussion}.

\section{Data preparation}\label{data_preparation}
\begin{figure*}
\centering
\includegraphics[width=1\textwidth]{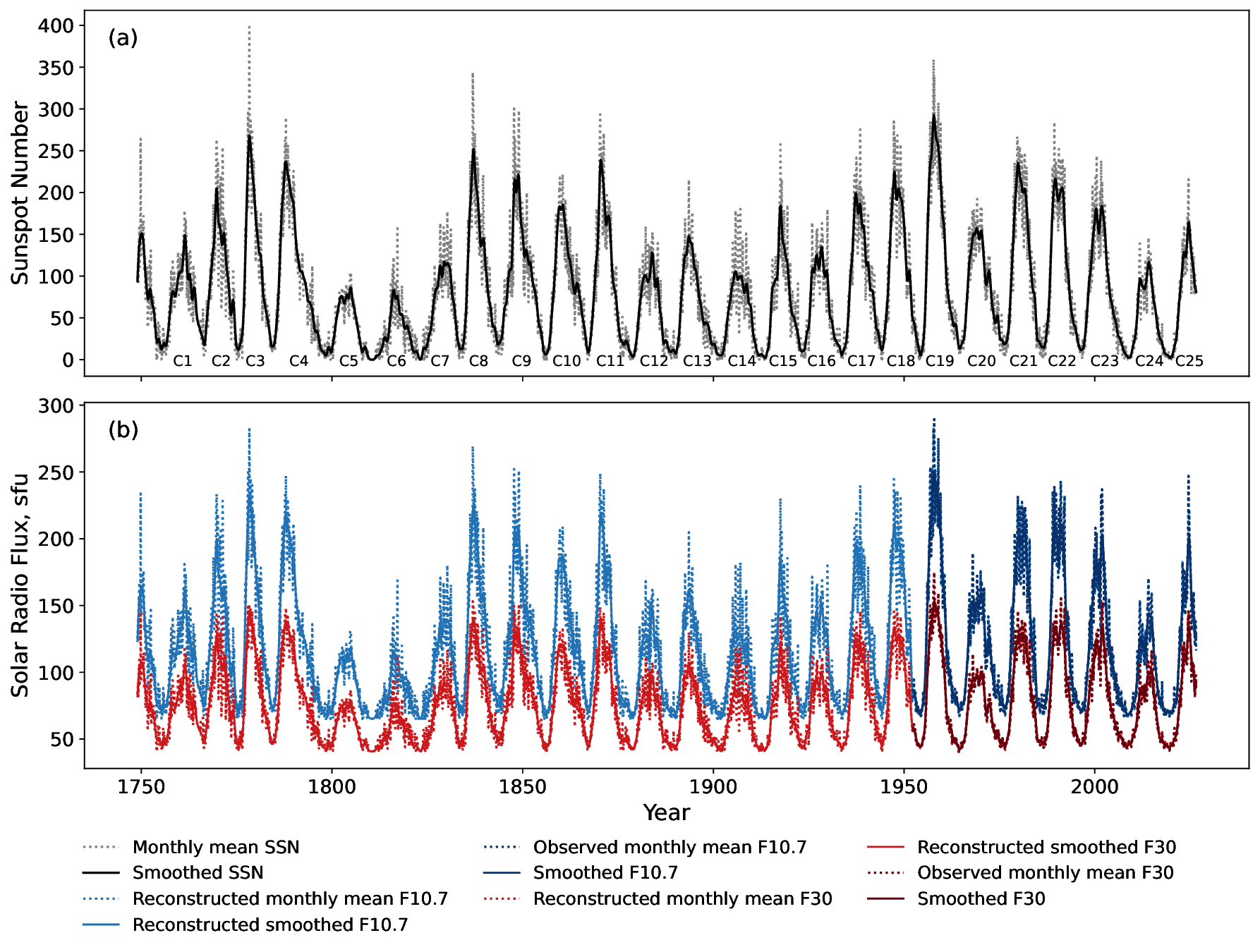}
\caption{Solar-activity records used in the analysis.
(a) Monthly (dotted gray) and 13-month smoothed (solid black) total sunspot numbers. C1-C25 denote solar cycles, defined by minima of the smoothed sunspot number. (b) Observed (dark blue and dark red) and reconstructed (light blue and light red) F10.7 and F30 radio flux records, including monthly values (dotted lines) and smoothed trends (solid lines).
\label{fig:SN_F107_F30_overview}}
\end{figure*}

We use monthly mean F10.7 and F30 radio flux from the Ottawa and Penticton observations and from the Toyokawa and Nobeyama facilities, respectively, together with the total sunspot number from SILSO. The international sunspot number record extends from January 1749 to August 2026 (the latest data considered in this study), whereas the F10.7 and F30 radio flux measurements are available only from November 1951, covering approximately 6.7 solar cycles. Since the M\&L method relies on the statistical properties of previous solar cycles, extending the radio flux records to earlier cycles provides a more representative basis for multi-year forecasting. Figure~\ref{fig:SN_F107_F30_overview} illustrates the solar activity records used in this study.  Panel (a) shows monthly mean sunspot numbers (gray dotted line) and their 13-month smoothed version (black solid line). Panel (b) shows the observed F10.7 and F30 radio flux records (dark blue and dark red dotted lines, respectively) and their 13-month smoothed versions (light blue and light red solid lines, respectively). The data are smoothed using the optimized 13-month smoothing technique described in \citet{Podladchikova2017} and \citet{Petrova2021}, which provides a balance between fidelity to observations and smoothness of the solar-cycle trend. The solar radio flux at wavelengths of 10.7~cm (F10.7) and 30~cm (F30) is expressed in solar flux units (sfu), where $1\,\mathrm{sfu}=10^{-22}\,\mathrm{W\,m^{-2}\,Hz^{-1}}$. 

We reconstruct the F10.7 and F30 radio-flux records before November 1951 using their relationships with sunspot numbers derived from third-order polynomial regressions
\begin{equation}
F_i=b_0+b_1R_i+b_2R_i^2+b_3R_i^3,
\label{eq:regression}
\end{equation}
Here, $R_i$ is the sunspot number and $F_i$ is the reconstructed radio flux. We fit the relationships separately between monthly mean sunspot number and monthly mean F10.7 and F30, and between smoothed sunspot number and smoothed F10.7 and F30, each over the interval where the respective flux record and sunspot number overlap. 
To investigate whether stronger smoothing provides a more stable representation of the solar-cycle trend for multi-year forecasting, we also prepare the corresponding reconstructed records using a 35-month smoothing window. The third-order polynomial allows for changes in the curvature of the relationship between sunspot numbers and solar radio flux across different levels of solar activity and provides the lowest reconstruction RMSE compared with linear and quadratic fits for both indices. The corresponding regression coefficients are listed in Table~\ref{tab:regression}. 
\begin{table}[htbp]\centering\small
\caption{Regression coefficients used for the reconstruction of historical F10.7 and F30 radio flux records. Coefficients are given for the monthly and for the 13- and 35-month smoothed series, together with the RMSE of each fit (in sfu)}\label{tab:regression}
\begin{tabular}{ccccccc}\toprule
Index & Series & $b_0$ & $b_1$ & $b_2$ & $b_3$ ($\times10^{-6}$) & RMSE (sfu)\\\midrule
F10.7 & Monthly & 65.1422 & 0.5483 & 0.0010 & $-$2.6059 & 10.68\\
F30 & Monthly & 41.1378 & 0.4264 & 0.0003 & $-$1.7678 & 9.09\\
F10.7 & 13-month smoothed & 65.3643 & 0.4947 & 0.0018 & $-$4.7610 & 6.64\\
F30 & 13-month smoothed & 40.8146 & 0.3984 & 0.0007 & $-$2.8317 & 6.55\\
F10.7 & 35-month smoothed & 65.4084 & 0.4897 & 0.0018 & $-$4.8584 & 6.55\\
F30 & 35-month smoothed & 40.8350 & 0.3966 & 0.0007 & $-$2.8634 & 6.48\\\bottomrule
\end{tabular}
\end{table}

The reconstructed F10.7 and F30 radio flux records, including monthly mean values and their 13-month smoothed series, are shown in Figure~\ref{fig:SN_F107_F30_overview} (panel b) as light blue and light red lines, respectively. The agreement between the reconstructed and observed flux values over the common interval is characterized by RMSE values of 11 sfu and 7 sfu for monthly and 13-month smoothed F10.7, respectively, and 9 sfu and 7 sfu for monthly and 13-month smoothed F30, respectively. Monthly records represent the observed radio flux activity and provide observations for the subsequent Kalman filter update, while smoothed records characterize the solar-cycle evolution used for M\&L forecasting.

\section{Initial multi-year McNish--Lincoln predictions }\label{initial_ML}

\begin{figure*}
\centering
\includegraphics[width=1\textwidth]{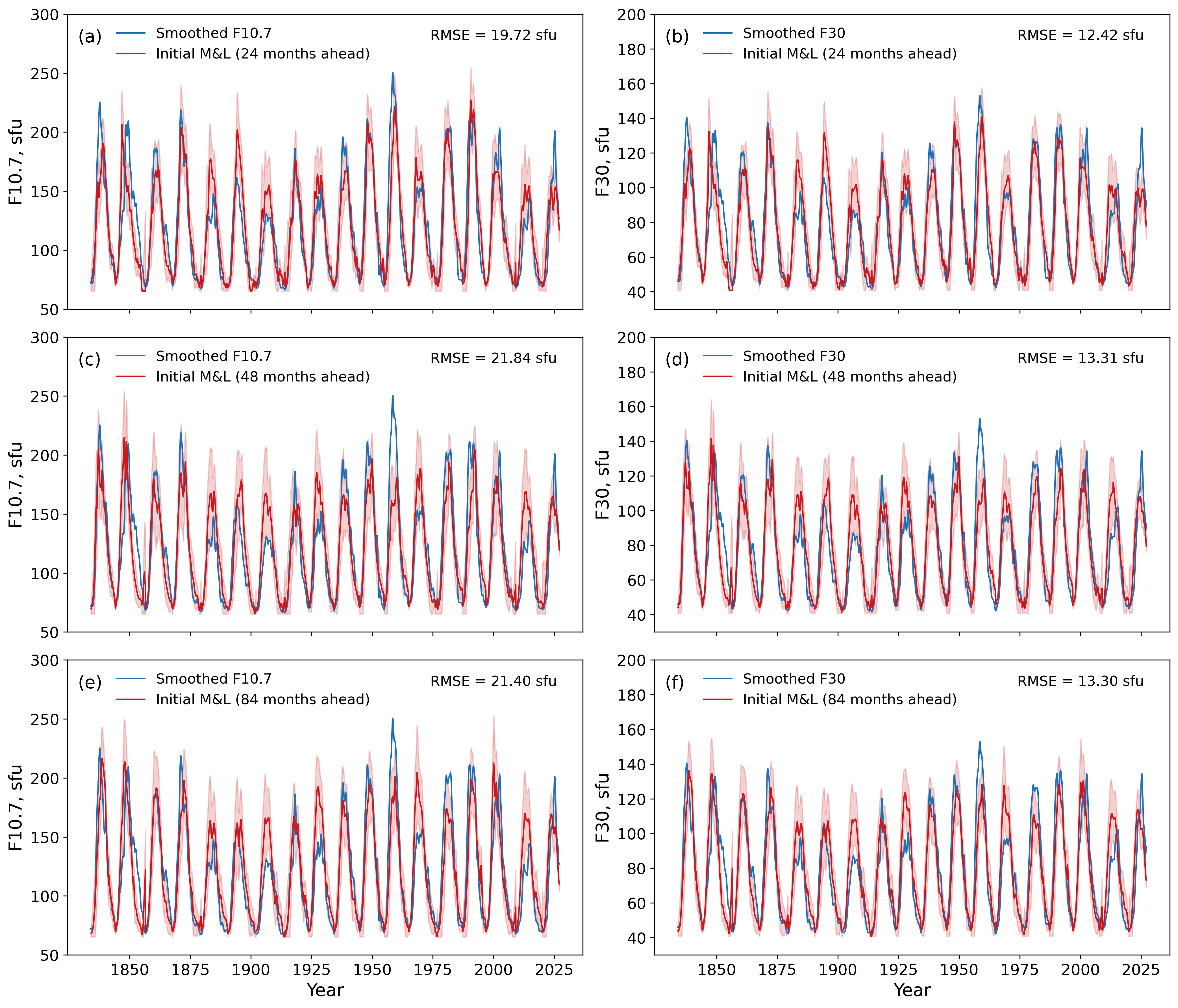}
\caption{Initial M\&L predictions of 13-month smoothed F10.7 (left column) and F30 (right column). The rows show predictions at 24, 48, and 84 months ahead. Blue curves give the smoothed radio flux and red curves the predicted values over cycles 8--25.  Shaded pink areas show $1\sigma$ prediction error defined by Eq~(\ref{eq_ML_analytical_error}).
\label{fig:Initial_ML_13_opt1}}
\end{figure*}

The long-term evolution of solar radio flux reflects the rise and decline of solar magnetic activity over the solar cycle. Although successive cycles share a common evolutionary pattern, they differ in amplitude and in the timing of their development. Our objective is to account for these differences using the observed evolution of the current cycle and the statistical properties of previous cycles. The M\&L method provides an empirical description of this evolution, allowing the future radio flux to be estimated from the current activity level and its relationship with the historical mean solar cycle.

The M\&L method uses a mean cycle constructed from reference cycles, i.e., past solar cycles aligned at their activity minima. Cycle boundaries are identified as the local minima of the smoothed flux, and each cycle's duration is the number of months between consecutive minima. The mean cycle duration is the average of these durations over the reference cycles, excluding cycle 25 because its duration is not yet observed, and the mean cycle is obtained by averaging, month by month, across the aligned reference cycles. The retrospective forecasts are evaluated over solar cycles 8--25 using leave-one-cycle-out validation. For each predicted cycle, the mean cycle is constructed from the remaining cycles, excluding the cycle being predicted. Earlier cycles are not included because the historical sunspot record before cycle 8 is less homogeneous.

Using M\&L, we predict the smoothed radio flux at a future month $m$ as
\begin{equation}
F_m=\overline{F}_m+k_{mi}(F_i-\overline{F}_i).
\label{eq:ml}
\end{equation}
Here, $F_i$ is the last available smoothed value at month $i$, and $\overline{F}_i$ and $\overline{F}_m$ are the mean-cycle values at the corresponding months. The mean cycle gives the initial estimate of future activity. The second term corrects this estimate according to the deviation of the activity level of the current cycle from the mean. The coefficient $k_{mi}$ is obtained by least squares \citep{Petrova2021}, excluding the cycle being predicted. This formulation accounts for differences between the current and historical mean solar cycles. The regression coefficient determines how the observed deviation from the mean cycle influences the prediction at different forecast horizons. Thus, the predicted radio flux reflects both the typical solar-cycle evolution and the observed activity level of the ongoing cycle.
 
We apply the same M\&L method to predictions from 1 to 84 months ahead for both indices (F10.7 and F30) and both smoothing windows (13 and 35 months). Since centered smoothing introduces a delay of 6 and 17 months for the 13- and 35-month windows, respectively, relative to the current month, the initial M\&L prediction must include this delay in the forecast horizon. Figure~\ref{fig:Initial_ML_13_opt1} shows examples of the initial M\&L predictions with lead times of 24 (top), 48 (middle), and 84 (bottom) months for the 13-month smoothed F10.7 (left column) and F30 (right column). The examples demonstrate that the M\&L method is able to reproduce the main evolution of the solar cycle even at multi-year lead times. The predicted curves preserve the timing of the rising and declining phases and provide a realistic estimate of the long-term activity level. The RMSE values vary with lead time and depend on the smoothing window, reflecting that the forecast is based on a statistical representation of a common solar-cycle shape and cannot fully capture the amplitude differences between individual cycles. The corresponding RMSE values range from 20 to 22 sfu for F10.7 and from 12 to 13 sfu for F30 and are indicated in each panel.

In addition to the RMSE values, Figure~\ref{fig:Initial_ML_13_opt1} shows the analytically derived $1\sigma$ prediction uncertainty, $\hat{\sigma}_m^{ML}$, of the initial M\&L method, obtained by propagating the uncertainties associated with the mean cycle and the current-cycle deviation. The prediction uncertainty is defined as
\begin{equation}\label{eq_ML_analytical_error}	\hat{\sigma}_{p}^{ML}=\sqrt{\frac{\left(\sigma_{m}^{2}-k_{mi}^{2}\sigma_{i}^{2}\right)\left(N_{c}-1\right)}{N_{c}-2}
	\left(1+\frac{1}{N_{c}}+\frac{\left(F_{i}^{c}-\bar{F}_{i}\right)^{2}}{\sigma_{i}^{2}\left(N_{c}-1\right)}\right)}.
\end{equation}
Here, $i$ and $m$ denote the prediction start and target months, respectively; $N_c$ is the number of cycles used to construct the mean cycle; $\sigma_i^2$ and $\sigma_m^2$ are the variances of the smoothed radio flux across these cycles at months $i$ and $m$; $F_i^c$ is the smoothed flux at month $i$ of cycle $c$, $\bar{F}_i$ is the corresponding mean-cycle value, and $k_{mi}$ is the M\&L regression coefficient. A detailed derivation is given in \citet{Petrova2021}; an equivalent formulation for sunspot-number predictions is presented in \citet{Clette2024}.

\begin{figure*}
\centering
\includegraphics[width=1\textwidth]{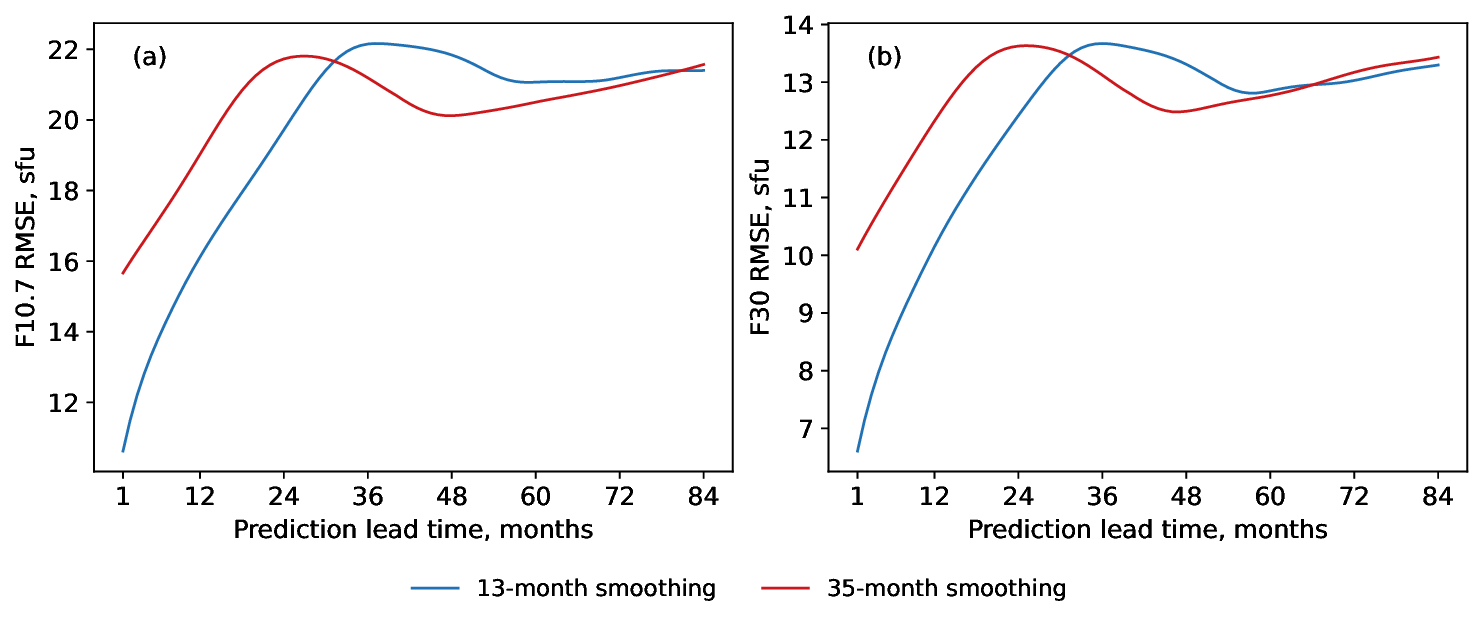}
\caption{RMSE of initial M\&L predictions from 1 to 84 months ahead for (a) F10.7 and (b) F30. Blue and red curves correspond to 13- and 35-month smoothing, respectively.  
\label{fig:RMSE_initial_M13_M35}}
\end{figure*}

Figure~\ref{fig:RMSE_initial_M13_M35} presents the RMSE of the initial M\&L predictions for solar cycles 8--25 over lead times from 1 to 84 months. For a smoothing window of 13 months, the RMSE ranges from approximately 11 to 22 sfu for F10.7 and from 7 to 14 sfu for F30. For a smoothing window of 35 months, the RMSE ranges from approximately 16 to 22 sfu for F10.7 and from 10 to 14 sfu for F30. The errors generally increase at shorter lead times and then vary more gradually over the multi-year interval. This behavior is particularly evident for F30 with 35-month smoothing, where the RMSE reaches approximately 14 sfu at a lead time of 24 months and remains within roughly 2 sfu of this value at longer lead times. The effect of smoothing depends on the forecast horizon, as the 13-month smoothed series provides lower errors at short lead times, whereas the 35-month smoothed series provides lower errors over part of the forecast range beyond approximately 30 months. The lower prediction errors obtained with 35-month smoothing over a substantial range of longer lead times suggest that stronger suppression of short-term variability provides a more stable representation of the underlying solar-cycle trend, despite the additional delay introduced by stronger smoothing. We calculated the RMSE of the initial M\&L forecasts separately for solar cycles 8--25, using the available portion of cycle 25. Table~\ref{tab:ML_RMSE_by_cycle} reports the minimum and maximum RMSE across cycles for each of nine lead times, both radio flux indices, and both smoothing windows. Either cycle 9 or cycle 19 had the largest RMSE in 34 of the 36 combinations examined. Cycle 9 was the longest completed cycle in our sample, and cycle 19 had the highest peak values of both F10.7 and F30. The two exceptions occurred at the 80-month lead time with 35-month smoothing, when cycle 8 had the largest RMSE for both indices. Under these conditions, the forecasts underestimated the flux around the maximum of cycle 8, which occurred about a year earlier relative to cycle onset than the median peak time for cycles 9--24. With 13-month smoothing, however, cycle 8 had the smallest RMSE for both indices at lead times of 60 and 70 months.

\begin{table}
\centering
\small
\setlength{\tabcolsep}{3pt}
\caption{Minimum and maximum RMSE (sfu) of the initial M\&L forecasts across solar cycles 8--25 at each selected lead time. Cycle numbers are given in parentheses. For cycle~25, the calculation uses the available portion of the cycle.}
\label{tab:ML_RMSE_by_cycle}
\begin{tabular}{@{}lcc*{9}{c}@{}}
\toprule
Index & Smoothing & RMSE & \multicolumn{9}{c}{Forecast lead time (months)} \\
\cmidrule(l){4-12}
      & (months) & & 1 & 10 & 20 & 30 & 40 & 50 & 60 & 70 & 80 \\
\midrule
F10.7 & 13 & Min. & 6.8 (14) & 8.1 (16) & 10.4 (11) & 12.6 (15) & 14.1 (17) & 12.4 (11) & 10.5 (8) & 10.2 (8) & 11.1 (17) \\
      &    & Max. & 16.0 (9) & 25.3 (9) & 33.9 (9) & 36.4 (19) & 40.5 (19) & 40.9 (19) & 37.8 (19) & 36.1 (9) & 34.1 (9) \\
\addlinespace
      & 35 & Min. & 9.0 (16) & 8.3 (15) & 14.1 (15) & 13.7 (17) & 12.2 (11) & 9.9 (11) & 10.8 (18) & 10.6 (17) & 11.1 (17) \\
      &    & Max. & 27.5 (9) & 34.4 (9) & 37.2 (19) & 40.1 (19) & 38.9 (19) & 35.0 (19) & 35.1 (9) & 32.0 (9) & 32.6 (8) \\
\midrule
F30   & 13 & Min. & 4.0 (21) & 5.4 (16) & 6.1 (20) & 6.9 (20) & 7.5 (20) & 7.7 (11) & 6.1 (8) & 6.7 (8) & 6.8 (17) \\
      &    & Max. & 9.9 (9) & 16.0 (9) & 21.3 (9) & 23.3 (9) & 23.3 (19) & 22.9 (19) & 21.1 (9) & 21.6 (9) & 20.9 (9) \\
\addlinespace
      & 35 & Min. & 5.4 (18) & 6.0 (15) & 6.8 (20) & 7.4 (20) & 7.4 (18) & 7.1 (18) & 7.6 (18) & 6.8 (17) & 7.2 (17) \\
      &    & Max. & 17.3 (9) & 21.4 (9) & 22.2 (9) & 22.4 (19) & 21.0 (19) & 20.3 (9) & 20.7 (9) & 19.5 (9) & 19.6 (8) \\
\bottomrule
\end{tabular}
\end{table}

\begin{figure*}
\centering
\includegraphics[width=1\textwidth]{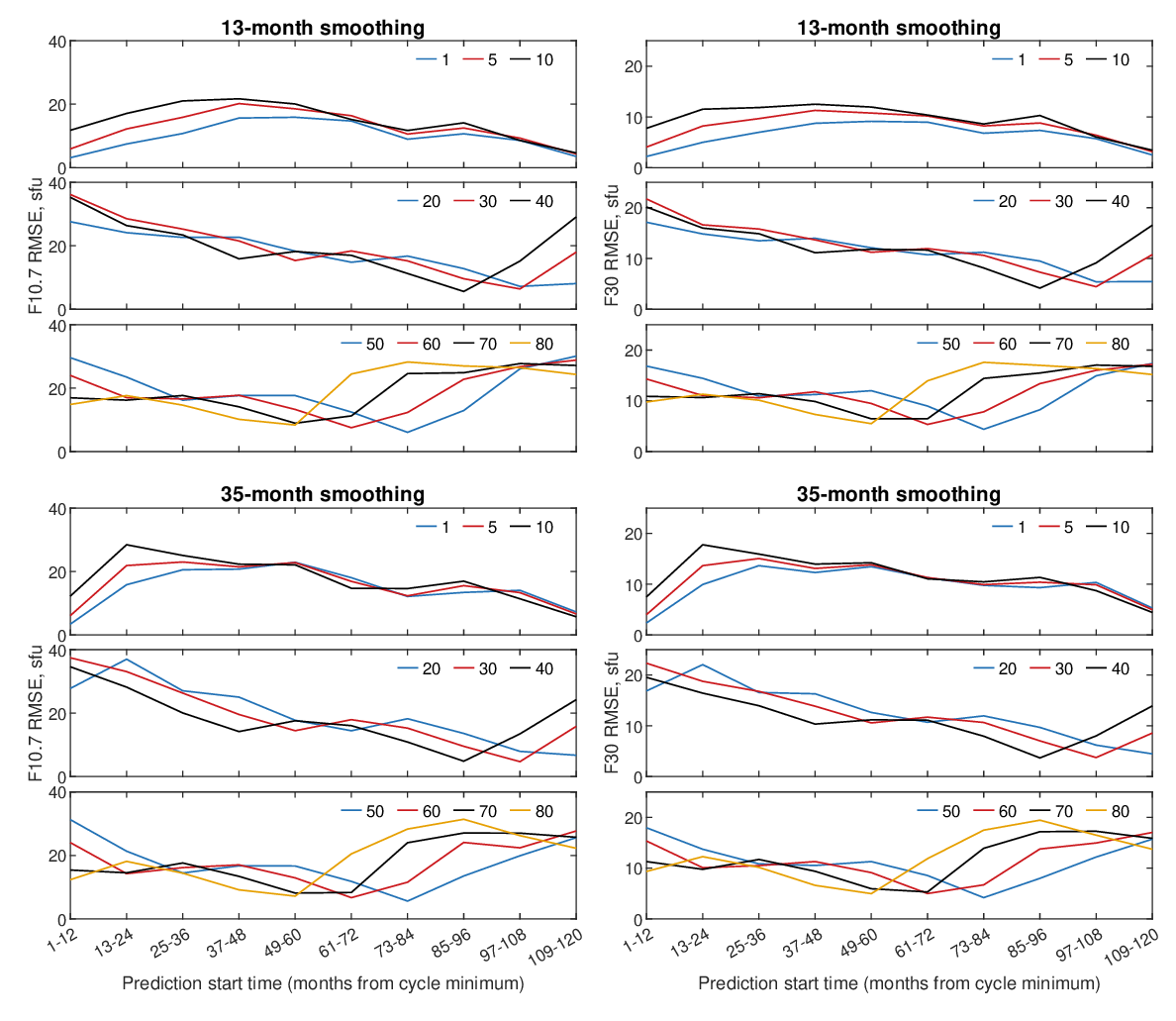}
\caption{Initial M\&L RMSE as a function of prediction start time for selected lead times of 1, 5, and 10--80 months in 10-month steps and two smoothing windows (13 and 35 months). F10.7 and F30 are shown in the left and right columns. The x-axis shows prediction start times grouped into 12-month intervals relative to the cycle minimum (e.g., ``1--12'' corresponds to forecasts issued during the first year after minimum). The prediction start time is the month at which the forecast is issued, and the lead time is counted from this month.} \label{fig:ML_phase_examples}
\end{figure*}

\begin{figure*}
\centering
\includegraphics[width=1\textwidth]{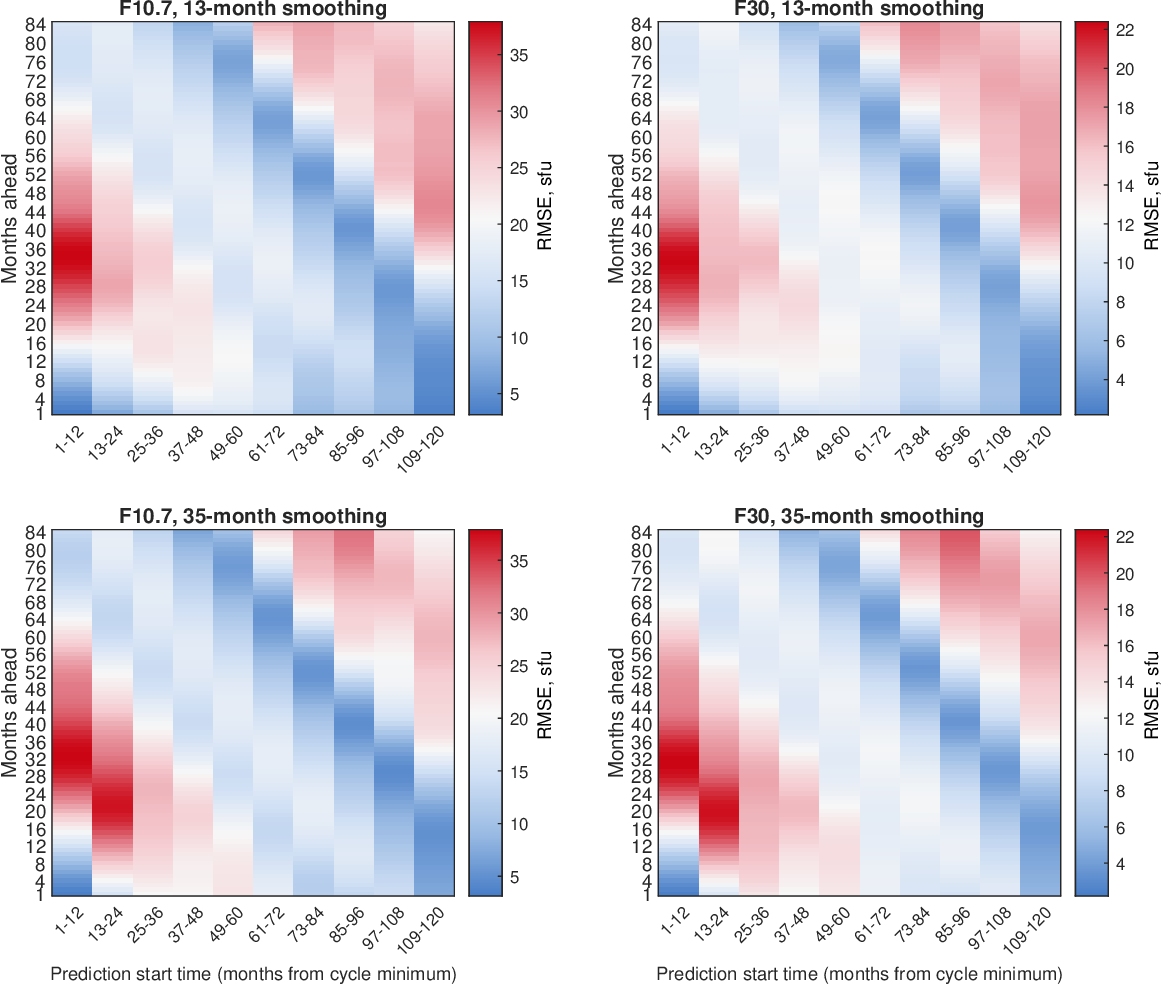}
\caption{Initial M\&L RMSE as a function of prediction start time ($x$-axis) and lead time ($y$-axis). The heatmaps show RMSE values over solar cycles 8-25 for lead times from 1 to 84 months. F10.7 and F30 are shown in the left and right columns, respectively, for the 13- and 35-month smoothing windows. 
The prediction start time and reference minimum are defined as in Figure 4.
\label{fig:RMSE_ML_start_heatmap}}
\end{figure*}

The initial M\&L predictions provide a multi-year reference whose accuracy depends not only on the forecast horizon but also on the phase of the solar cycle at which the prediction is initiated. To investigate this dependence, we analyze the RMSE as a function of the prediction start time, grouped into 12-month intervals relative to the minimum of the solar cycle in which the forecast is issued. The prediction start time is the current month, i.e., the month at which the forecast is issued, and the lead time is counted from this month. For the initial M\&L, the last available smoothed value lags the current month by 6 months (13-month window) or 17 months (35-month window), so the forecast uses that value and this delay is included in the forecast horizon. Figure~\ref{fig:ML_phase_examples} shows this dependence for lead times of 1, 5, and 10--80 months in 10-month steps, using 13- and 35-month smoothing windows. We examine the results for short-, intermediate-, and long-term predictions, with numerical examples given for 13-month smoothing.

For short-term predictions (1--10 months ahead), the lowest errors occur when forecasts are initiated shortly after solar minimum or during the late declining phase, while starts during the rising phase give larger errors. For example, at a 10-month lead time, starting at 109--120 months rather than 37--48 months after minimum reduces the F10.7 RMSE from 22 to approximately 5 sfu and the F30 RMSE from 12 to 3 sfu. Thus, even for short-term predictions, the accuracy strongly depends on the starting phase.

For intermediate lead times (20--40 months), the lowest errors shift toward forecasts initiated during the declining phase. At a 40-month lead time, starting at 85--96 months rather than within the first 12 months after minimum reduces the F10.7 RMSE from 35 to 6 sfu and the F30 RMSE from 20 to 4 sfu. However, starting later, at 109--120 months, increases the errors again to 29 and 16 sfu, respectively. The favorable starting phase therefore depends on the forecast horizon.

For long-term predictions (50--80 months), the lowest errors occur progressively earlier in the solar cycle as the lead time increases. For F10.7, the most favorable starting interval shifts from 73--84 months after minimum for a 50-month forecast to 61--72 months for a 60-month forecast and 49--60 months for 70--80-month forecasts. The corresponding minimum RMSEs are approximately 6, 8, and 9 sfu, respectively. F30 follows the same pattern, with minimum errors of approximately 4--6 sfu. The dependence on the prediction start time is also evident at longer lead times. For an 80-month forecast, the F10.7 RMSE increases from approximately 9 to 27 sfu for predictions initiated 49--60 and 85--96 months after minimum, respectively.

These results show that solar radio flux predictability depends not only on the forecast horizon but also on the starting and target phases of the solar cycle. In particular, the lowest errors at long lead times generally occur when the predicted month falls near the late declining phase or the subsequent solar minimum. Thus, low forecast errors can be achieved even at multi-year horizons, provided that the prediction starts at a favorable phase of the solar cycle.

The effect of smoothing on forecast accuracy depends on both lead time and prediction start time. For short-term predictions initiated during the rising phase, 13-month smoothing generally gives lower errors. For a 10-month forecast initiated 13--24 months after minimum, the F10.7 RMSE is approximately 18 sfu with 13-month smoothing, compared with 28 sfu with 35-month smoothing. The corresponding F30 errors are 12 and 18 sfu.

At longer lead times, 35-month smoothing can reduce prediction errors, depending on the starting phase. For a 40-month forecast initiated 109--120 months after minimum, stronger smoothing reduces the F10.7 RMSE from 29 to 24 sfu and the F30 RMSE from 16 to 14 sfu. However, for an 80-month forecast initiated 85--96 months after minimum, the F10.7 RMSE increases from 27 to 31 sfu.

These results demonstrate that the choice of smoothing window depends on both the forecast horizon and the starting phase of the solar cycle. While 13-month smoothing provides lower errors for short-term predictions, 35-month smoothing can improve multi-year forecasts, particularly at lead times of 36--60 months (Figure~\ref{fig:RMSE_initial_M13_M35}), depending on the prediction start time (Figure~\ref{fig:ML_phase_examples}).

Figure~\ref{fig:RMSE_ML_start_heatmap} brings these results together over the full range of lead times from 1 to 84 months and all considered prediction start intervals. Both indices show broad regions of comparatively low RMSE, including at multi-year lead times. The largest errors (about 35--38 sfu for F10.7 and 20--22 sfu for F30) occur for predictions initiated within the first 12 months after the cycle minimum and extending approximately 30--40 months after. Errors also increase when the forecast is initiated late in the current cycle and the predicted month falls within the high-activity phase of the following cycle. 

The low-RMSE regions extend to lead times of 50--80 months, where the favorable starts shift from approximately 73--84 to 49--60 months after minimum, with RMSEs of about 5--10 sfu for F10.7 and 3--6 sfu for F30. These predictions approximately reach the late declining phase or the subsequent minimum. These results show that M\&L predictions can have low RMSE even at lead times of 50--80 months, with the lowest errors occurring when the predicted month falls near the subsequent solar minimum. For practical forecasting, these results identify the horizons over which M\&L provides the lowest errors for a given prediction start time. Starts near the cycle minimum favor short forecasts, while starts around the maximum and during the declining phase can support low-error predictions extending several years ahead. The 35-month smoothing window additionally reduces the overall RMSE at lead times of approximately 36--60 months (Figure~\ref{fig:RMSE_initial_M13_M35}), with the phase maps showing where this advantage occurs. Together, these results support multi-year radio-flux forecasting over a seven-year range and provide an empirical basis for assessing the expected accuracy at each starting phase and forecast horizon. These results show that solar radio flux predictability depends not only on the forecast horizon but also on the phase of the solar cycle. Forecast errors can remain relatively low even at lead times of 50--80 months when the predicted activity falls near the subsequent solar minimum.

However, the initial M\&L prediction relies on the mean-cycle representation and a delayed smoothed value, without explicitly updating the current solar-cycle state as new observations become available. In the next section, we therefore introduce an adaptive Kalman filter approach that incorporates monthly observations to estimate the current solar-cycle state and uses this updated state to refine the initial M\&L prediction.

\section{Updating multi-year predictions with an adaptive Kalman filter}\label{KFML}
\begin{figure*}
	\centering
	\includegraphics[width=0.8\textwidth]{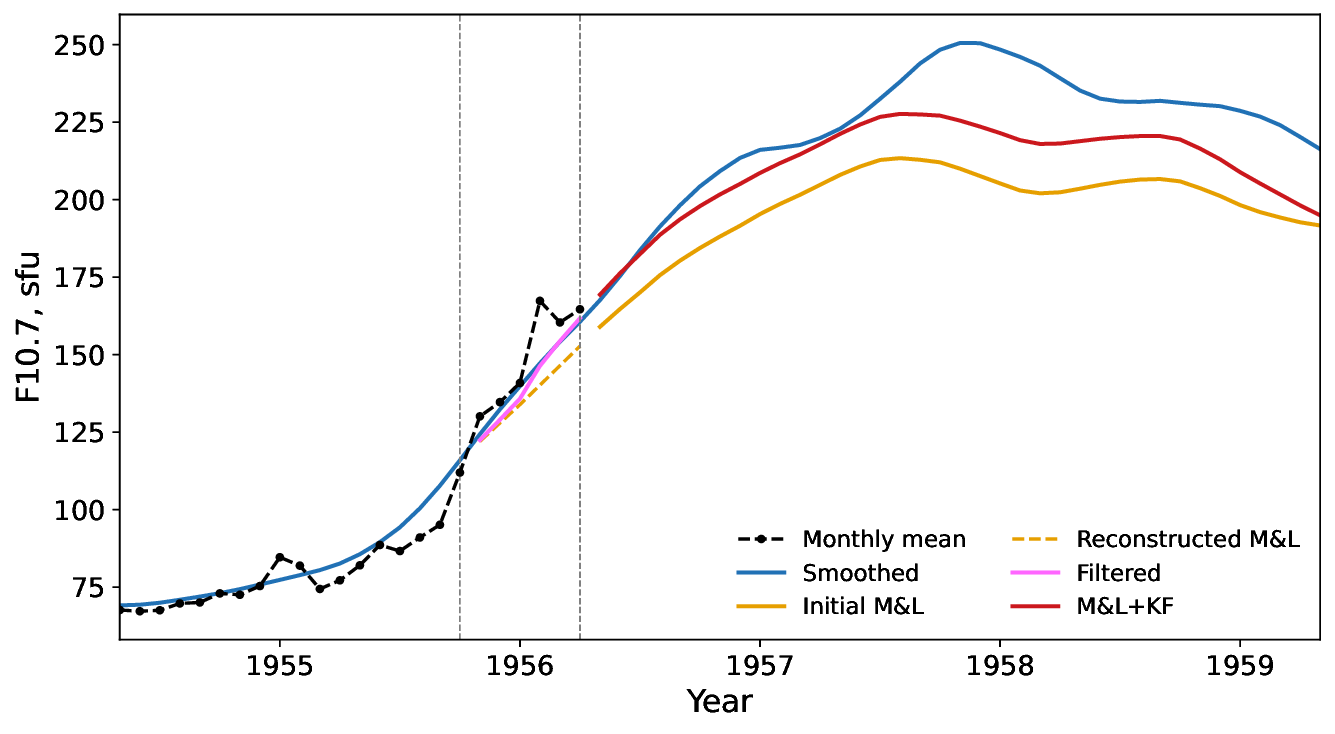}
	\caption{Illustration of the adaptive Kalman filter update for F10.7 during cycle 19, with predictions up to 37 months ahead. The black dashed line with circles shows the monthly mean observations, and the blue line shows the smoothed radio flux, including values available only retrospectively. The vertical gray dashed lines mark the time of the last available 13-month smoothed value (left) and the current month (right). The orange dashed line shows the initial M\&L estimate over the interval between these dates, and the magenta line shows the Kalman-filtered estimate obtained by assimilating the monthly observations. Beyond the current month, the orange solid and red lines show the initial and updated M\&L predictions, respectively. The updated prediction starts from the filtered estimate of the current smoothed flux.
  \label{fig:Illustration_plot}}
\end{figure*}

\begin{figure*}
	\centering
	\includegraphics[width=1\textwidth]{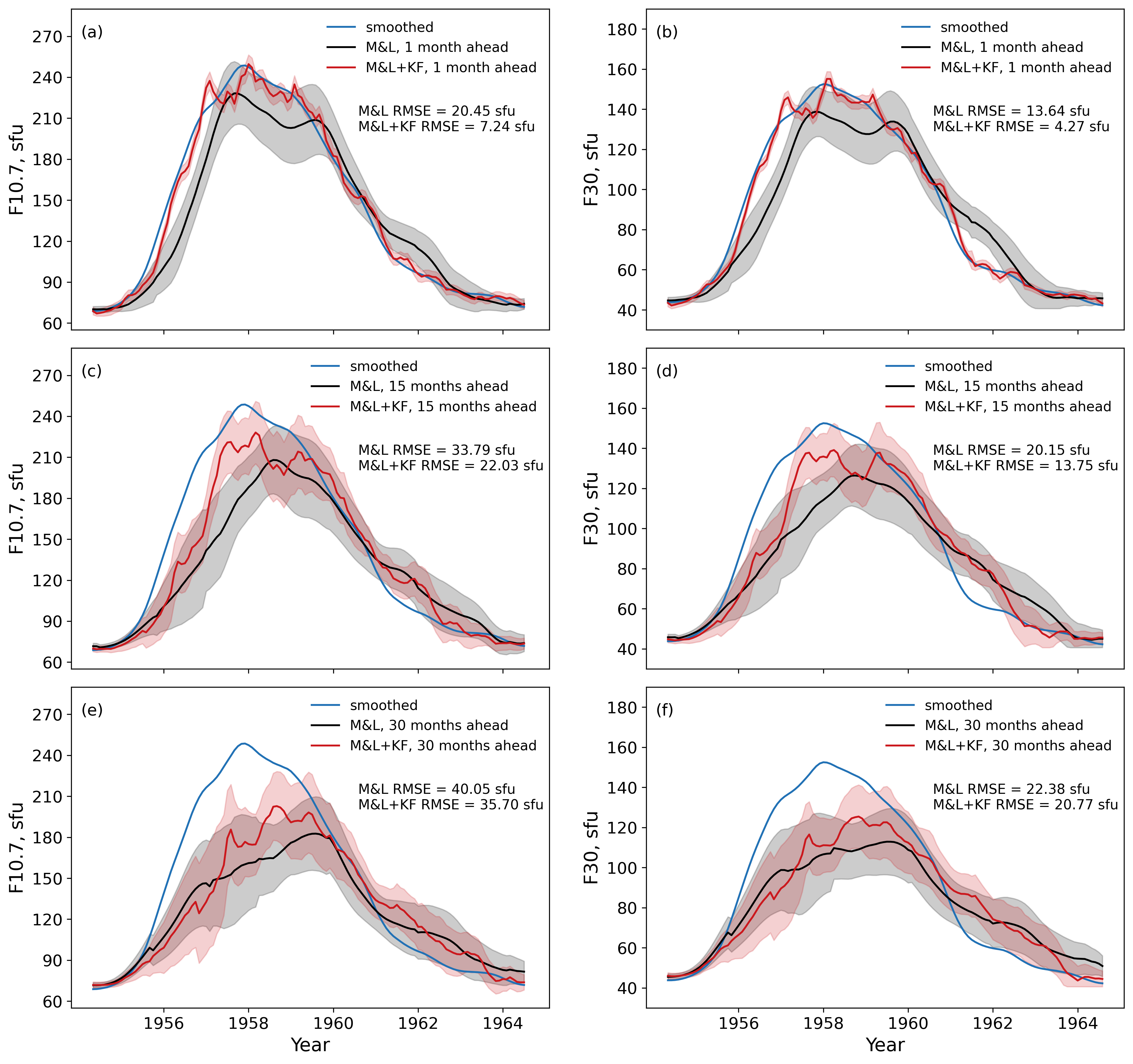}
	\caption{Initial (M\&L) and Kalman-updated (M\&L+KF) predictions of 35-month smoothed F10.7 (left column) and F30 (right column) during solar cycle 19. The rows show predictions at lead times of 1, 15, and 30 months, respectively. Blue curves show the smoothed radio flux, while black and red curves represent the initial and updated predictions. Gray and pink shaded bands indicate the corresponding analytical $1\sigma$ prediction uncertainties (Equations~(\ref{eq_ML_analytical_error})~and~(\ref{eq:KFML_uncertainty})).
		\label{fig:Truth_ML_MLKF_C19}}
\end{figure*}

The initial M\&L prediction describes the expected evolution of the solar-cycle activity, while the latest monthly radio-flux observations provide information about its actual development. By combining the model prediction with these observations, we reconstruct the current smoothed radio flux, distinguishing changes in the underlying solar-cycle trend from short-term variability.
To sequentially update the expected solar-cycle evolution with new observations, we formulate the problem in state-space form. The smoothed radio flux $F_j$ describes the underlying solar-cycle activity at month $j$, while the monthly mean flux $F_j^{\mathrm{obs}}$ provides the corresponding observation. The state and observation equations are
\begin{equation}
\begin{aligned}
F_j=\Phi_{j,j-1}F_{j-1}+w_j,\\
F_j^{\mathrm{obs}}=F_j+\eta_j,
\end{aligned}
\label{eq:state_space_model}
\end{equation}
where $w_j$ and $\eta_j$ denote zero-mean uncorrelated model and measurement noise, respectively; $j = 1,...,6$ for 13-month smoothing, and $j = 1,...,17$ for 35-month smoothing. The model noise accounts for uncertainties in the initial M\&L predictions, while the measurement noise represents the short-term variability of the monthly observations relative to the smoothed flux.

The state evolution is prescribed by the initial M\&L prediction, shown by the orange dashed line over the delayed interval in Figure~\ref{fig:Illustration_plot}. 
Thus, the model follows the expected evolution of solar-cycle activity inferred from previous cycles and the observed activity level of the current cycle. The transition coefficient describes the relative change in the predicted radio flux between consecutive months, allowing the state-space model to reproduce this evolution.
Denoting the initial prediction by $F_j^{\mathrm{ML}}$, we define the state-transition coefficient $\Phi_{j,j-1}$, which in the present one-dimensional formulation reduces to a scalar and is therefore referred to as the transition coefficient:
\begin{equation}
\Phi_{j,j-1}
=
\frac{F_j^{\mathrm{ML}}}{F_{j-1}^{\mathrm{ML}}},
\label{eq:KF_transition}
\end{equation}
where $F_0^{\mathrm{ML}}$ is the last available smoothed value. At each month, we first propagate the state estimate according to this state-transition coefficient:
\begin{equation}
	\begin{aligned}
		\widehat{F}_{j,j-1}
		&=\Phi_{j,j-1}\widehat{F}_{j-1,j-1},\\
		\sigma_{j,j-1}^{2}
		&=\Phi_{j,j-1}^{2}\sigma_{j-1,j-1}^{2}
		+\sigma_{w,j}^{2},
	\end{aligned}
	\label{eq:KF_prediction}
\end{equation}
where $\sigma_{j,j}^{2}$ is the error variance of the filtered estimate $\widehat{F}_{j,j}$, and $\sigma_{w,j}^{2}$ is the model noise variance. We then incorporate the monthly observation $F_j^{\mathrm{obs}}$ to obtain the filtered estimate $\widehat{F}_{j,j}$. First, we compute the Kalman gain, which determines the weight assigned to the new observation when correcting the predicted state:
\begin{equation}
	K_j=\frac{\sigma_{j,j-1}^{2}}{\sigma_{j,j-1}^{2}+\sigma_{\eta,j}^{2}}.
	\label{eq:KF_gain}
\end{equation}
Here, $\sigma_{j,j-1}^{2}$ is the error variance of the predicted estimate $\widehat{F}_{j,j-1}$, and $\sigma_{\eta,j}^{2}$ is the measurement noise variance. Next, we correct the predicted estimate and update its error variance:
\begin{equation}
	\begin{aligned}
		\widehat{F}_{j,j} &= \widehat{F}_{j,j-1}+K_j\left(F_j^{\mathrm{obs}}-\widehat{F}_{j,j-1}\right),\\
		\sigma_{j,j}^{2} &= (1-K_j)\,\sigma_{j,j-1}^{2}.
	\end{aligned}
	\label{eq:KF_update}
\end{equation}

In Figure~\ref{fig:Illustration_plot}, the monthly observations indicate a faster increase in activity than the initial M\&L prediction. Their assimilation raises the filtered estimate (magenta line), bringing it closer to the smoothed flux (blue line). This example illustrates how recent observations reveal departures of the ongoing solar cycle from its initially predicted evolution. Incorporating these observations allows us to reconstruct the current activity level and correct the subsequent prediction of the rising phase and solar maximum.

Reliable identification of $\sigma_{w,j}^{2}$ and $\sigma_{\eta,j}^{2}$ is essential for the Kalman filter, since these quantities determine the balance between the model prediction and the observations. Incorrect specification of model and observation error statistics can lead to substantial estimation errors and even failure of the data assimilation algorithm \citep{Podladchikova2014_RB_1,Podladchikova2014_RB_2}. We therefore develop an updated noise statistics identification procedure for solar radio-flux forecasting, building on \citet{Podladchikova2012}. This update provides a more reliable specification of the model and measurement uncertainties, leading to improved filtered estimates of the smoothed flux. The procedure and its modifications are described in Appendix~\ref{appendix_NSI}. Table~\ref{tab:rmse_results_current_point} compares the filtered estimate at the final assimilation step with the initial M\&L estimate of the same current-month smoothed flux, obtained by predicting 6 or 17 months from the last available smoothed value. The improvement is quantified by the relative
reduction of the RMSE,
\begin{equation}
  \Delta\mathrm{RMSE} =
  \dfrac{\mathrm{RMSE}_{\mathrm{M\&L}} - \mathrm{RMSE}_{\mathrm{M\&L+KF}}}
        {\mathrm{RMSE}_{\mathrm{M\&L}}} \times 100\%,
  \label{eq:drmse}
\end{equation}
so that positive values indicate a lower error for M\&L+KF.
The filter reduces RMSE by 39--41\% for 13-month smoothing and by 62\%
for 35-month smoothing. The filter reduces RMSE by 39--41\% for 13-month smoothing and by 62\% for 35-month smoothing. For both windows, the filtered estimates have RMSEs of approximately 6 sfu for F10.7 and 3.5--3.7 sfu for F30. The filtered estimate at the current month provides a new starting value for M\&L, from which we generate the updated multi-year predictions. Because the filtered estimate is available at the current month, the starting point of the updated forecast coincides with the prediction start time defined in Section~\ref{initial_ML}, whereas for the initial M\&L it lags the start time by 6 or 17 months. The initial and updated predictions are therefore compared at the same start time and lead time. In Figure~\ref{fig:Illustration_plot}, the updated prediction (red line) follows the subsequent rise more closely and reduces the underestimation of the cycle maximum compared with the initial prediction (orange solid line).

\begin{table}
	\centering
	\caption{RMSE of current-time reconstruction of the smoothed F10.7 and F30 indices, cycles 8--25. M\&L denotes the reconstruction obtained from the M\&L prediction made 6 months (13-month window) or 17 months (35-month window) ahead from the last available smoothed value, while M\&L+KF denotes the Kalman-filtered estimate of the same point. $\Delta$RMSE is the relative RMSE reduction of M\&L+KF with respect to M\&L.}
	\label{tab:rmse_results_current_point}
	\begin{tabular}{llcccc}
		\toprule
		\multirow{2}{*}{\shortstack{Smoothing \\ Window}} & \multirow{2}{*}{\shortstack{Reconstruction \\ method}} & \multicolumn{2}{c}{F10.7 cm} & \multicolumn{2}{c}{F30 cm} \\
		\cmidrule(lr){3-4} \cmidrule(lr){5-6}
		& & RMSE & $\Delta$RMSE & RMSE & $\Delta$RMSE \\
		\midrule
		\multirow{2}{*}{13 months} & M\&L      & 9.66  &     & 6.03 &     \\
		                           & M\&L+KF   & 5.94  & 39\%  & 3.53 & 41\%  \\
		\midrule
		\multirow{2}{*}{35 months} & M\&L      & 15.38 &     & 9.90 &     \\
		                           & M\&L+KF   & 5.88  & 62\%  & 3.72 & 62\%  \\
		\bottomrule
	\end{tabular}\\
\end{table}

Figure~\ref{fig:Truth_ML_MLKF_C19} compares the initial (black lines) and updated (red lines) predictions of 35-month smoothed F10.7 and F30 against the smoothed flux (blue lines) during cycle 19, for lead times of 1, 15, and 30 months. At a one-month lead time, the F10.7 RMSE decreases from 20 to 7 sfu and the F30 RMSE from 14 to 4 sfu (panels a and b). The improvement remains substantial up to 15-month lead time (panels c and d), with RMSE decreasing from 34 to 22 sfu and from 20 to 14 sfu, respectively. At 30 months, the reduction is smaller, and both methods underestimate the cycle maximum (panels e and f). Thus, this example demonstrates that updating the current smoothed-flux estimate can improve subsequent predictions at both short and multi-year lead times.

The shaded bands in Figure~\ref{fig:Truth_ML_MLKF_C19} represent the analytically derived $1\sigma$ prediction uncertainties for the initial M\&L and updated M\&L+KF predictions, given by Equation~(\ref{eq_ML_analytical_error}) and the following expression, respectively
\begin{equation}
	\widehat{\sigma}_{p}^{\mathrm{KFML}}
	=
	\sqrt{
	\frac{(\sigma_m^2-k_{mi}^2\sigma_i^2)(N_c-1)}{N_c-2}
	\left(
	1+\frac{1}{N_c}
	+\frac{(\widehat{F}_{i}^{c}-\overline{F}_i)^2}
	{\sigma_i^2(N_c-1)}
	\right)
	+k_{mi}^2\sigma_{i,i}^2}
	\label{eq:KFML_uncertainty}
\end{equation}
Here, the quantities are defined as in Equation~(\ref{eq_ML_analytical_error}), however, here we use the estimated $\widehat{F}_{i}^{c}$ with the Kalman filter instead of the available smoothed radio flux ${F}_{i}^{c}$. $\sigma_{i,i}^2$ is the filtration error variance at the final assimilation step, given by Equation~(\ref{eq:KF_update}). The term $k_{mi}^2\sigma_{i,i}^2$ propagates this uncertainty in the filtered starting value to the updated prediction. In these examples, the analytical prediction uncertainties are generally smaller for M\&L+KF than for the initial M\&L predictions. Thus, the uncertainty in the reconstructed current activity level contributes to the forecast uncertainty, with its influence depending on the prediction horizon through the M\&L regression coefficient.

\begin{figure*}
	\centering
	\includegraphics[width=1\textwidth]{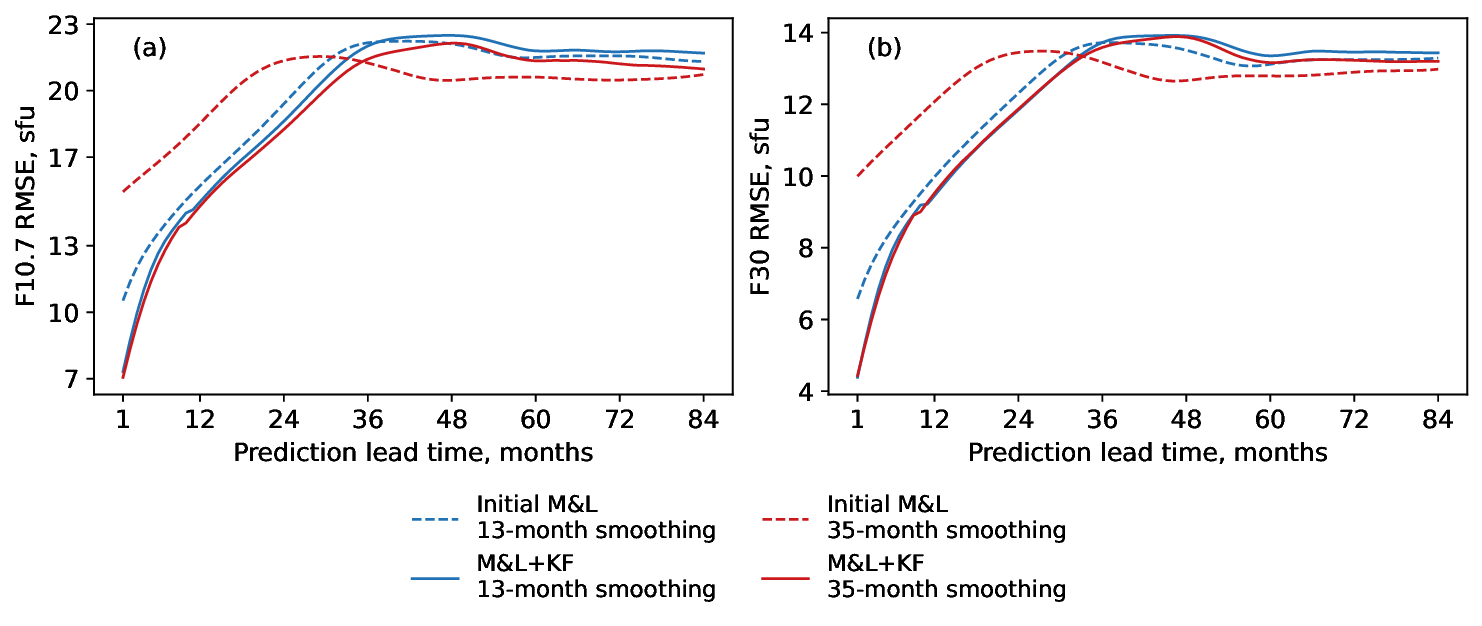}
	\caption{RMSE of the initial M\&L predictions (dashed lines) and Kalman-updated predictions (solid lines) over solar cycles 9--25 for lead times of 1--84 months for F10.7 (panel a) and F30 (panel b). Blue and red curves correspond to 13- and 35-month smoothing.
		\label{fig:RMSE_ML_KFML_M13_M35}}
\end{figure*}

To determine how representative these improvements are across solar cycles and lead times, Figure~\ref{fig:RMSE_ML_KFML_M13_M35} compares the RMSE of the initial M\&L predictions (dashed lines) and the updated predictions (solid lines) for lead times of 1--84 months. For consistency, the RMSE values in this comparison are evaluated over cycles 9--25 to account for the shifts introduced by the update procedure. Panels (a) and (b) show F10.7 and F30, respectively, with blue and red curves corresponding to 13- and 35-month smoothing.

The largest improvement occurs at short lead times, particularly for 35-month smoothing. At a one-month lead time, the update reduces the F10.7 RMSE from approximately 15 to 7 sfu and the F30 RMSE from 10 to 4 sfu. For 13-month smoothing, the corresponding reductions are from approximately 10 to 7 sfu and from 7 to 4 sfu. These results highlight the complementary roles of stronger smoothing and the Kalman update. The 35-month window provides a more stable representation of the long-term solar-cycle trend, while the assimilation of recent observations compensates for the longer delay in estimating the current activity level. As a result, the update substantially reduces short-term prediction errors and brings the two smoothing windows to a similar level of accuracy. For both smoothing windows, the update reduces RMSE on average up to lead times of approximately 30--35 months, with the largest gains at short horizons. At longer lead times, the overall errors are comparable with those of the initial M\&L predictions. However, improvements can persist at longer horizons for particular prediction start times, as discussed below in Figures~\ref{fig:KFML_phase_examples} and~\ref{fig:KFML_heatmap}.

\begin{figure*}
		\centering
		\includegraphics[width=1\textwidth]{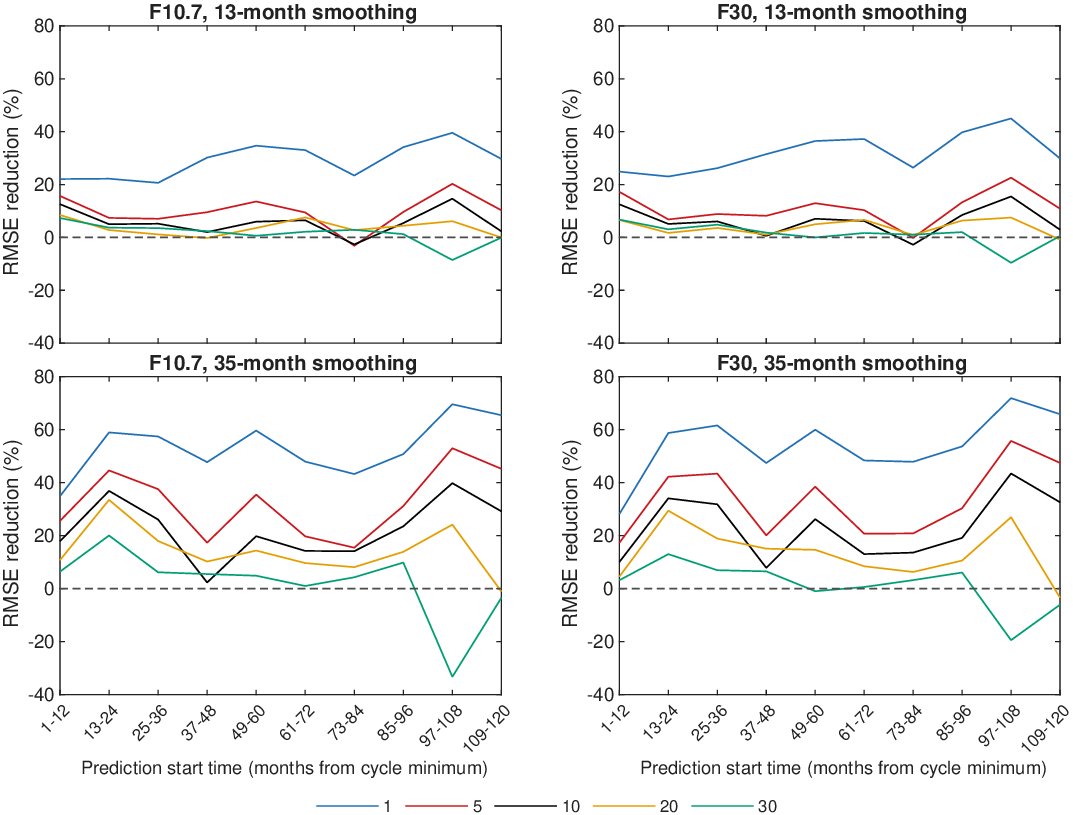}
		\caption{Relative RMSE reduction from the Kalman update as a function of prediction start time for selected lead times of 1, 5, 10, 20, and 30 months. F10.7 and F30 are shown in the left and right columns, with 13- and 35-month smoothing in the upper and lower rows, respectively. Positive values indicate lower RMSE for M\&L+KF relative to the initial M\&L predictions; negative values indicate higher RMSE.
			\label{fig:KFML_phase_examples}}
\end{figure*}
   
\begin{figure*}
		\centering
		\includegraphics[width=1\textwidth]{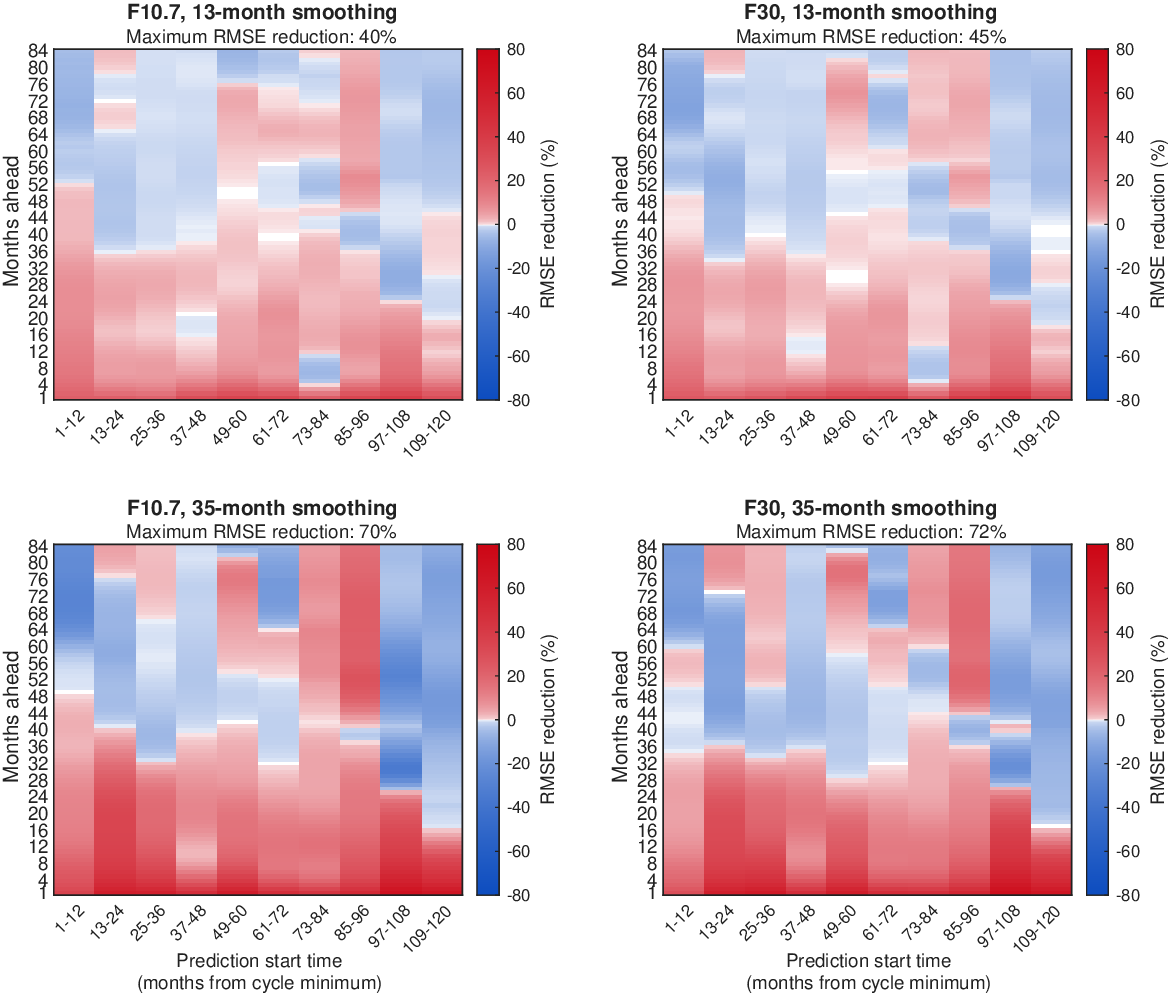}
		\caption{Relative RMSE reduction from the Kalman update as a function of prediction start time and lead time over the full 1--84-month forecast range. F10.7 and F30 are shown in the left and right columns, with 13- and 35-month smoothing in the upper and lower rows, respectively. Positive values (red) indicate a reduction in RMSE achieved by the Kalman update, whereas negative values (blue) indicate an increase in RMSE relative to the initial M\&L predictions.
			\label{fig:KFML_heatmap}}
\end{figure*}


Figure~\ref{fig:KFML_heatmap} extends the selected examples in Figure~\ref{fig:KFML_phase_examples} to the full range of lead times from 1 to 84 months. For both indices, the update reduces RMSE over a broad range of prediction start times, with larger improvements for 35-month smoothing. For many starting intervals, the benefit extends to approximately 30--40 months ahead and can persist to about 45 months ahead for forecasts initiated within the first 12 months after minimum. For late-cycle starts, this range becomes shorter: approximately 24 months ahead for starts at 97--108 months after minimum and 15--20 months ahead for starts at 109--120 months. The heat maps also reveal additional regions of improvement at longer lead times, beyond the approximately 35-month range identified from the overall RMSE in Figure~\ref{fig:RMSE_ML_KFML_M13_M35}. With 35-month smoothing, forecasts initiated 85--96 months after minimum benefit from the update at lead times of approximately 45--84 months, with RMSE reductions of about 20--30\% at 48--60 months ahead for both indices. For starts at 49--60 months after minimum, reductions of approximately 10--20\% occur at lead times of 70--80 months. Similar regions are also present with 13-month smoothing, demonstrating that both smoothing windows support improved multi-year predictions when forecasts are initiated at favorable phases. 

These results demonstrate that incorporating recent observations provides a more accurate estimate of the current solar-cycle activity and improves subsequent predictions. This improvement arises because the Kalman update allows M\&L predictions to restart from the estimated current smoothed activity rather than from a value delayed by 6 or 17 months. This changes both the starting deviation from the mean cycle and the corresponding M\&L regression coefficient. Since differences in solar-cycle strength may persist over several years, the updated starting point can remain informative even at longer forecast horizons, although its benefit depends on the prediction start time and lead time.

For practical forecasting, these results favor the Kalman-updated predictions at lead times of approximately 1--35 months for most starting phases. For forecasts initiated 97--120 months after minimum, the preferred range is shorter, approximately 15--24 months ahead. At multi-year horizons, the update is particularly useful with 35-month smoothing for starts at 85--96 months after minimum and lead times of 45--84 months, and for starts at 49--60 months with lead times of 70--80 months. These intervals provide empirical guidance for selecting the updated predictions beyond the range of improvement in the overall RMSE. Such phase-dependent improvements are relevant for long-term thermospheric modeling and satellite mission planning, where reliable predictions of future solar activity are required.

\section{Forecasting with hemispheric solar activity information}\label{Hemispheric}

\begin{figure*}
\centering
\includegraphics[width=1\textwidth]{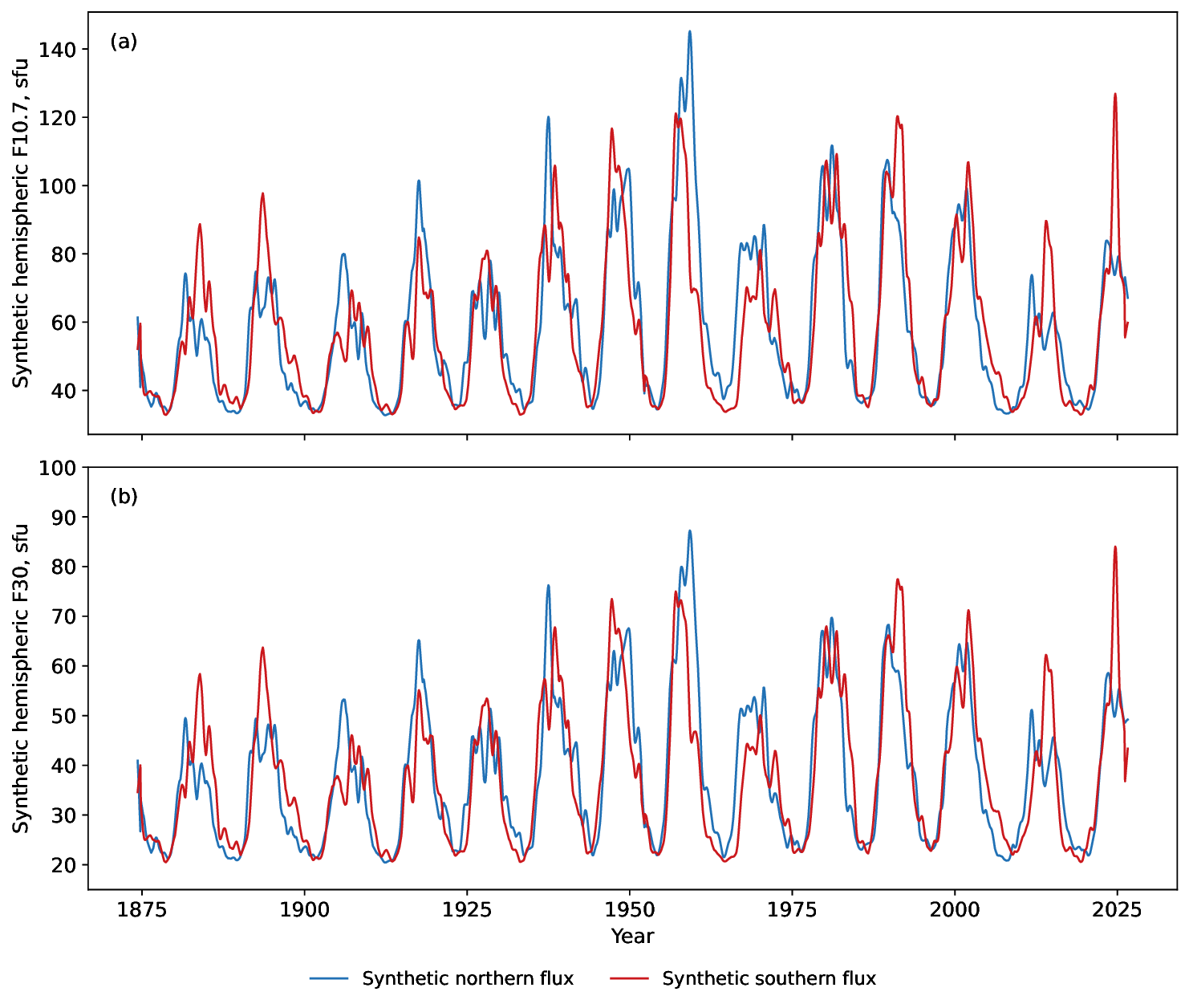}
\caption{Synthetic 13-month smoothed northern (blue) and southern (red) components of (a) F10.7 and (b) F30 over solar cycles 12--25, constructed from the total smoothed radio flux using time-dependent hemispheric sunspot-number fractions.
\label{fig:hemispheric_recon}}
\end{figure*}

\begin{figure*}
\centering
\includegraphics[width=1\textwidth]{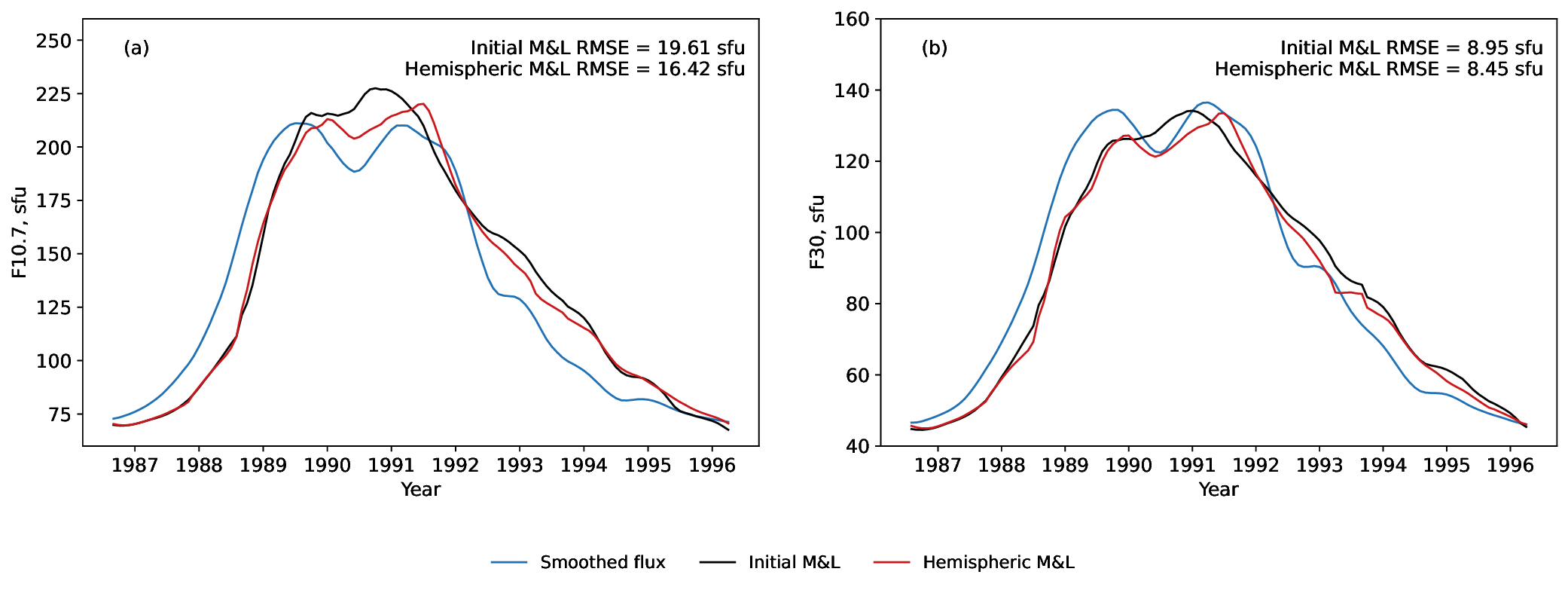}
\caption{Initial and hemispheric M\&L predictions of 13-month smoothed F10.7 (left column) and F30 (right column) during solar cycle 22 at a lead time of 18 months. Blue curves show the smoothed radio flux, while black and red curves represent the predictions using total flux and predictions using synthetic hemispheric data, respectively.
\label{fig:hem_pred_C22_m18_M13}}
\end{figure*}

\begin{figure*}
\centering
\includegraphics[width=1\textwidth]{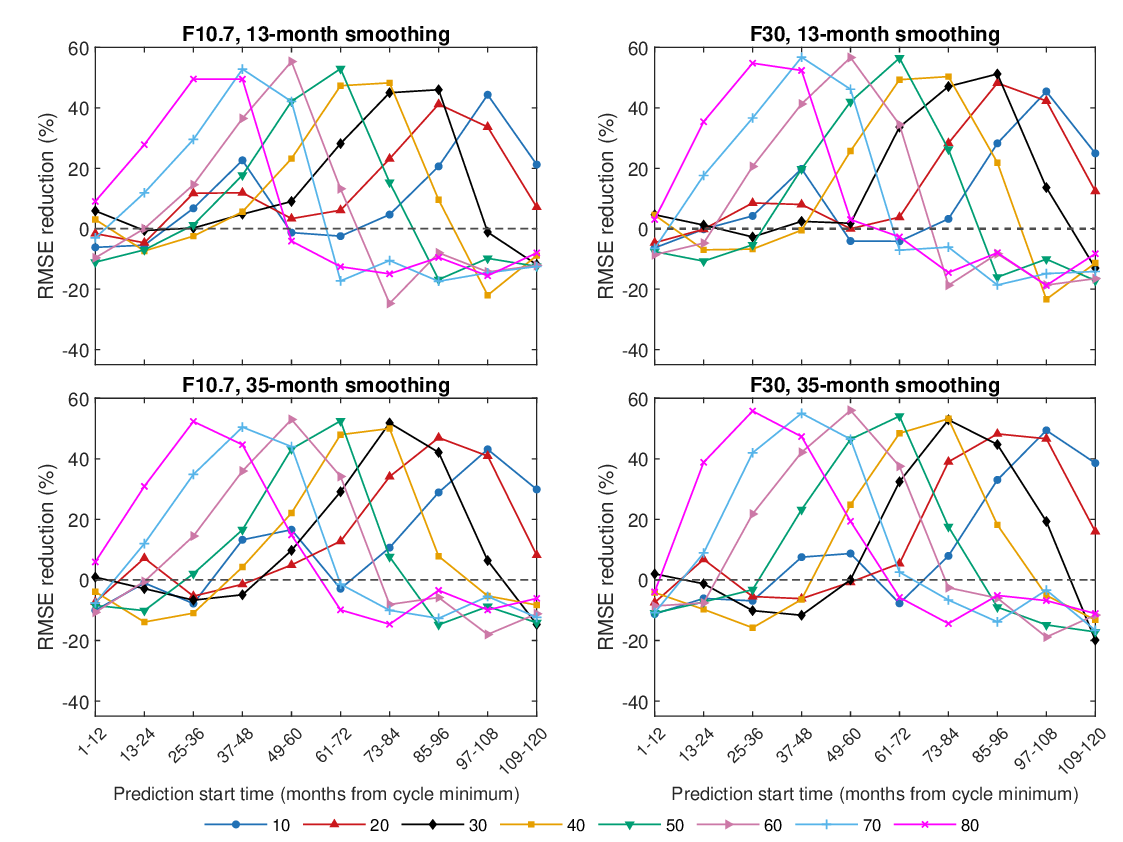}
\caption{Relative RMSE reduction from hemispheric M\&L forecasting as a function of prediction start time for selected lead times of 10--80 months in 10-month steps. F10.7 and F30 are shown in the left and right columns, with 13- and 35-month smoothing in the
upper and lower rows, respectively. Positive values indicate lower RMSE for hemispheric M\&L relative to the initial M\&L predictions;
negative values indicate higher RMSE.
\label{fig:examples_hemispheric}}
\end{figure*}

\begin{figure*}
\centering
\includegraphics[width=1\textwidth]{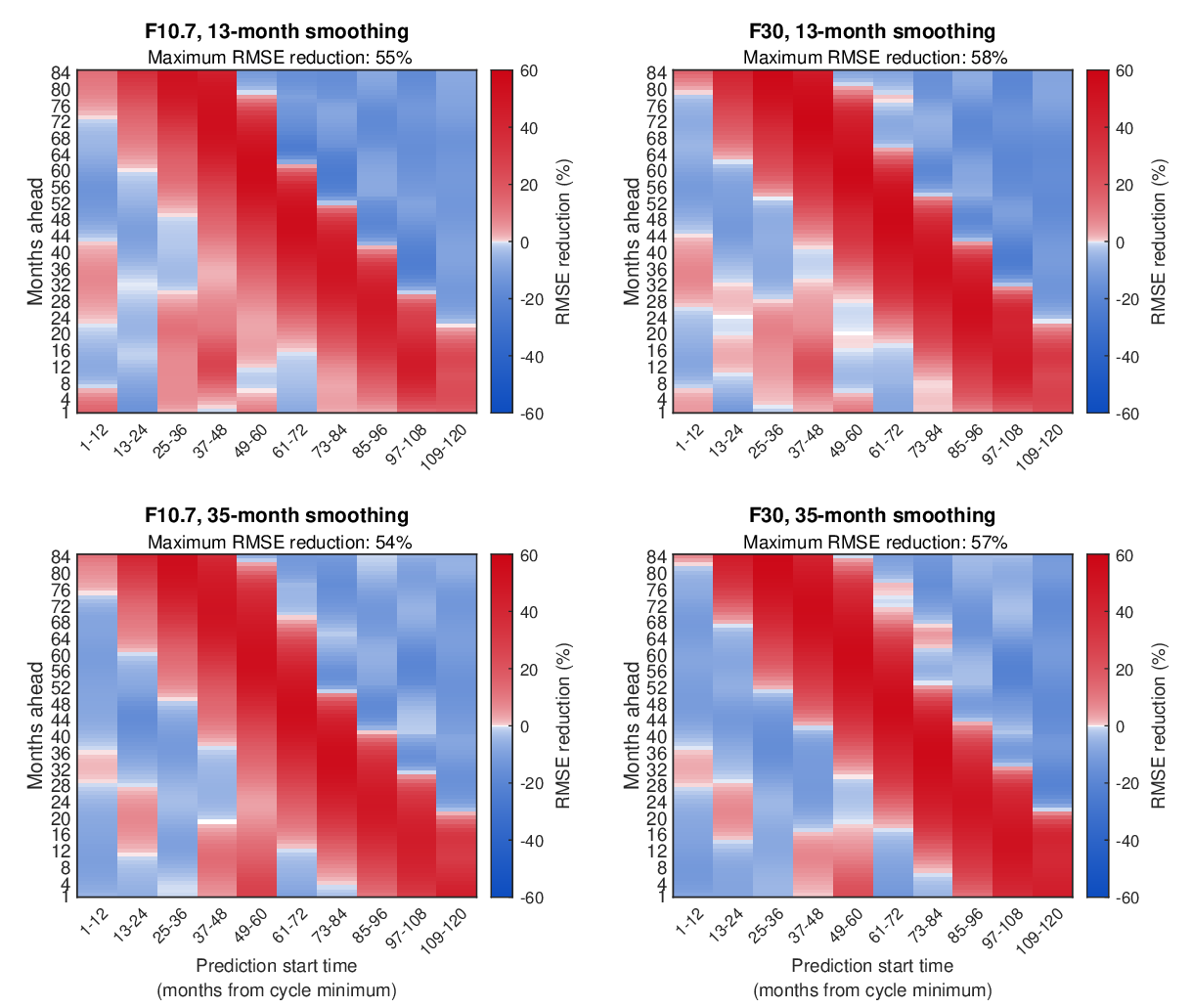}
\caption{Relative RMSE reduction from hemispheric M\&L forecasting as a function of prediction start time and lead time over the full 1--84-month forecast range. 
F10.7 and F30 are shown in the left and right columns, with 13- and 35-month smoothing in the upper and lower rows, respectively. Red regions indicate lower RMSE for hemispheric M\&L relative to the initial M\&L predictions, while blue regions indicate higher RMSE. \label{fig:heatmap_hemispheric}}
\end{figure*}

The forecasts considered so far use total solar activity. We now investigate whether accounting for the separate evolution of the northern and southern hemispheres can improve multi-year radio-flux predictions. Previous studies have demonstrated the predictive value of hemispheric solar activity. \citet{Veronig2021} found a stronger relationship between sunspot growth rates and cycle amplitude when the hemispheres were considered separately, and \citet{Podladchikova2022} showed that this improves cycle-amplitude predictions. Here, we extend this idea to radio-flux forecasting by constructing synthetic hemispheric flux series that retain the differences in timing and amplitude between northern and southern activity.

We use the smoothed total sunspot-number record and the smoothed total radio-flux record, including the historical flux reconstruction described in Section~\ref{data_preparation}. Monthly hemispheric sunspot numbers from 1874--2020 are taken from the extended catalogue of \citet{Veronig2021} and extended beyond 2020 using the WDC--SILSO monthly hemispheric series, version 2.0\footnote{\url{https://www.sidc.be/SILSO/datafiles}}. We first apply 13- and 35-month smoothing to the monthly hemispheric sunspot numbers, then we derive hemispheric flux components using the smoothed hemispheric sunspot numbers together with the regression coefficients obtained when fitting smoothed total sunspot number to smoothed total flux (Section~\ref{data_preparation}). For each index (F10.7 and F30), let $b_0, b_1, b_2, b_3$ denote the coefficients of that cubic regression.
The same regression coefficients are used for both hemispheres, so that the hemispheric partition remains consistent with the total flux--sunspot relation. We evaluate the activity-related part of the regression separately for the smoothed northern and southern hemispheric sunspot numbers $SN_j^N$ and $SN_j^S$ as
\begin{equation}
\begin{aligned}
A_j^N &= b_1 SN_j^N + b_2 \left(SN_j^N\right)^2 + b_3 \left(SN_j^N\right)^3,\\
A_j^S &= b_1 SN_j^S + b_2 \left(SN_j^S\right)^2 + b_3 \left(SN_j^S\right)^3.
\end{aligned}
\label{eq:hemispheric_activity}
\end{equation}
The constant term $b_{0}$ is omitted at this stage because it represents the background radio-flux contribution and is therefore not assigned according to hemispheric sunspot activity.

The hemispheric fractions are then defined as the relative contribution of each hemisphere to this activity-related term,
\begin{equation}
f_j^N = \frac{A_j^N}{A_j^N+A_j^S}, \qquad f_j^S = \frac{A_j^S}{A_j^N+A_j^S}.
\label{eq:hemispheric_fractions}
\end{equation}
The total smoothed flux $F^{T}_j$ is then split into hemispheric components by assigning half of the baseline term $b_0$ to each hemisphere and dividing the remaining activity-driven flux according to the fractions:
\begin{equation}
F_j^N = \frac{b_0}{2} + f_j^N \left(F^{T}_j - b_0\right), \qquad F_j^S = \frac{b_0}{2} + f_j^S \left(F^{T}_j - b_0\right).
\label{eq:hemispheric_flux}
\end{equation}
This construction preserves the total radio flux exactly, $F_j^N+F_j^S=F_j^T$, while the hemispheric sunspot numbers determine only how its activity-related component is partitioned between the two hemispheres. This procedure is applied separately for F10.7 and F30, using each index's respective regression coefficients.

Figure~\ref{fig:hemispheric_recon} shows the resulting series for 13-month smoothed hemispheric radio flux series over solar cycles 12–25. The northern (blue) and southern (red) components exhibit distinct peak amplitudes and timing, retaining hemispheric differences that are combined in the total flux. These sunspot-derived proxies provide a basis for testing whether the separate hemispheric evolution improves predictions of the total radio flux.

We incorporate hemispheric information into M\&L by constructing separate northern and southern mean cycles, predicting each hemispheric component, and then summing the two predictions to obtain the total flux prediction.  We evaluate the hemispheric M\&L predictions over solar cycles 13–25. As an illustrative example, Figure~\ref{fig:hem_pred_C22_m18_M13} shows an 18-month-ahead prediction of F10.7 and F30 for cycle 22 using the 13-month smoothed series. The smoothed flux exhibits a pronounced double-peaked maximum, with a decrease between the two peaks corresponding to the Gnevyshev gap. While the initial M\&L prediction largely smooths over this structure, the hemispheric M\&L prediction better reproduces the double-peaked maximum and the intervening gap. This improvement is also reflected in the RMSE over the full cycle, which decreases from 19.61 to 16.42 sfu for F10.7 and from 8.95 to 8.45 sfu for F30. Improved representation of the peak region is particularly relevant for applications sensitive to the high-activity phase of the solar cycle.

Figure~\ref{fig:examples_hemispheric} shows the RMSE reductions of hemispheric M\&L relative to the initial M\&L predictions for lead times of 10--80 months in 10-month steps. The approach shows a systematic shift of the largest improvements toward earlier prediction starts as the lead time increases. For example, the reduction at 40 months ahead peaks at starts 73--84 months after minimum, reaching approximately 50\% for both indices. At 60 months ahead, the peak shifts to starts at 49--60 months, with reductions reaching about 50--55\%, while at 80 months ahead it shifts further to starts at 25–36 months, where the reductions again reach about 50\% or more. The shift largely follows the geometry of the forecast, as reaching a similar target interval requires an earlier start as the lead time increases. The examples above place the predicted month approximately 110--120 months after minimum, suggesting that the strongest gains are associated with predicting the late declining phase and approach to the subsequent minimum. Retaining hemispheric differences may help describe this evolution more accurately, although the shift alone does not establish its physical cause. Practically, the hemispheric approach supports improved multi-year predictions from starts around maximum and during the declining phase. Such improvements are relevant to orbital-lifetime and re-entry assessments, which depend on the evolution of thermospheric drag over the solar cycle \citep{Petrova2021}.

Figure~\ref{fig:heatmap_hemispheric} extends the comparison to all lead times from 1 to 84 months.
Hemispheric M\&L shows a broad band of improvement that shifts toward earlier prediction starts as the forecast horizon increases. The maximum RMSE reductions reach 55\% for F10.7 and 58\% for F30 with 13-month smoothing, and 54\% and 57\%, respectively, with 35-month smoothing. Substantial improvements occur particularly at multi-year lead times. The maps provide concrete guidance for forecasting. For predictions 70--84 months ahead,  hemispheric M\&L is favorable when initiated approximately 25–48 months after minimum, with reductions reaching 40--50\% within this region. For starts at 49--60 months, the largest gains occur approximately 50--70 months ahead, whereas starts at 61--84 months favor horizons of approximately 30--50 months. These intervals follow the same shift identified in Figure~\ref{fig:examples_hemispheric}, broadly corresponding to predictions toward the late declining phase and subsequent minimum.  The method also provides substantial short-horizon improvements for late-cycle starts. For predictions initiated 97–108 months after minimum, RMSE reductions of roughly 30–50\% are obtained at lead times of about 10–20 months, depending on the index and smoothing window.

These results demonstrate the value of hemispheric information for multi-year radio-flux forecasting. By retaining differences in the amplitude and timing of northern and southern activity, the synthetic hemispheric series preserve information about the asymmetric evolution of solar magnetic activity that is obscured in the total index. 
Hemispheric M\&L substantially improves predictions over identifiable ranges of starting phases and forecast horizons.

\section{Discussion and Conclusions}\label{Discussion}
The main results of this study can be summarized in three main advances. First, we extended the McNish--Lincoln (M\&L) method to multi-year forecasting of the F10.7 and F30 solar radio flux indices, with lead times of 1--84 months, and quantified how forecast accuracy depends on both lead time and the phase of the solar cycle at which the forecast is issued. Second, we developed an adaptive Kalman filter with an updated identification of noise statistics. The filter estimates the smoothed radio flux for the current month, providing a starting point for multi-year predictions without the delay introduced by centered smoothing and improving the accuracy of subsequent forecasts. Third, we developed a hemispheric M\&L formulation in which the northern and southern components are predicted separately and then summed to obtain the total radio flux prediction. This formulation demonstrates that the separate hemispheric evolution provides additional predictive information and substantially improves multi-year forecasts.

Multi-year forecasting requires a statistically representative description of the solar cycle. Since direct radio flux measurements cover only about 6.7 cycles, we reconstructed F10.7 and F30 records back to cycle~1 from sunspot numbers using third-order polynomial regressions fitted to monthly mean and smoothed series separately. The reconstruction agrees with observations with RMSE values of 10.7 and 6.7~sfu for monthly and 13-month smoothed F10.7, and 9.1 and 6.5~sfu for monthly and 13-month smoothed F30. These records for cycles 8--25 are used to construct the mean cycle and assess forecast uncertainties. We also demonstrated that 35-month smoothing provides lower initial M\&L prediction errors than 13-month smoothing at lead times of approximately 32–80 months for F10.7 and 32–66 months for F30, suggesting that stronger smoothing provides a more stable representation of the long-term solar-cycle trend, despite the additional smoothing delay.

The initial M\&L predictions demonstrate that the mean-cycle representation retains predictive skill over multi-year horizons, extending from 1 to 84 months ahead. Over cycles 8--25, the RMSE varies between about 11 and 22~sfu for F10.7 and between 7 and 14~sfu for F30 with 13-month smoothing, and between 16 and 22~sfu and 10 and 14~sfu, respectively, with 35-month smoothing. The error generally increases during the first part of the forecast horizon, reaching a plateau of about 22~sfu for F10.7 after roughly 36 months and remaining within about 1~sfu of this level over the subsequent multi-year interval, indicating that the accuracy of a six-year forecast is not systematically worse than that of a three-year forecast.

Forecast accuracy depends not only on the lead time but also on the solar-cycle phase at which the forecast is initiated. For forecasts up to about one year ahead (1--10 months), the lowest errors are obtained when the prediction is initiated during the late declining phase of the cycle, whereas forecasts started shortly after solar minimum are less accurate because the amplitude of the upcoming maximum is still uncertain. For intermediate lead times of about 20--40 months, the most favorable starting phase shifts toward the post-maximum declining phase of the cycle. For longer forecasts of 50--80 months, the optimal starting interval moves further backward, toward the middle of the cycle. For example, with a 10-month lead time, the F10.7 RMSE decreases from about 22~sfu for forecasts initiated 37--48 months after minimum to about 5~sfu for starts at 109--120 months. For 70--80-month forecasts, the minimum errors are obtained for starts around the middle of the cycle (49--60 months after minimum). This behavior reflects the limited information available near solar minimum, when the upcoming maximum is not yet constrained by observations and the prediction relies more strongly on the statistical mean-cycle shape used in the M\&L method.

The adaptive Kalman filter with noise statistics identification compensates for the delay introduced by centered smoothing by assimilating the latest monthly observations and reconstructing the current-time smoothed radio flux. This update improves the initial M\&L predictions not only at short horizons but also over multi-year timescales, reducing RMSE by up to 40--45\% for 13-month smoothing and up to 70\% for 35-month smoothing. The improvement is most pronounced up to lead times of about 30--35 months, for which the Kalman-updated predictions are generally preferred across most starting phases. For forecasts initiated late in the cycle (97--120 months after minimum), the improvement is mainly limited to shorter horizons of about 15--24 months. At multi-year horizons, the Kalman correction remains advantageous for 35-month smoothing, particularly for starts at 85--96 months after minimum (45--84 months ahead) and at 49--60 months after minimum (70--80 months ahead). By compensating for the delay introduced by stronger smoothing, the filter extends the applicability of M\&L forecasting beyond the medium-term range. The proposed Kalman filter technique provides a universal post-processing framework for improving forecasts whenever observations and an initial predicted trajectory are available.

Using synthetic hemispheric radio flux series reconstructed from the extended North-South hemispheric sunspot-number record, we incorporated the separate evolution of northern and southern activity into the M\&L method. The hemispheric analysis demonstrates that retaining hemispheric evolution provides additional predictive information beyond the total (full-disk) solar index.
Hemispheric M\&L improves multi-year radio flux forecasts, with maximum RMSE reductions reaching 55\% for F10.7 and 58\% for F30, depending on the starting phase and forecast horizon. Hemispheric forecasting is most advantageous for long-horizon predictions initiated during the rising phase, around solar maximum, and during the declining phase, with the favorable lead time becoming shorter for later starting phases. For example, forecasts initiated 25--48 months after minimum benefit most strongly at 70--84 months ahead, whereas starts at 49--60 and 61--84 months after minimum provide improvements at approximately 50--70 and 30--50 months ahead, respectively. 
Beyond the reduction in overall RMSE, hemispheric information can improve the representation of complex structure around solar maximum by the better reproduction of the double-peaked maximum and the intervening Gnevyshev gap. These results demonstrate that hemispheric asymmetries contain predictive information that is lost when solar activity is represented only by the full-disk index.

The results obtained in this study open several directions for further development and operational application of multi-year solar radio flux forecasting. Future work will include the integration of hemispheric forecasting with the adaptive Kalman update and application of the developed techniques to other solar and geomagnetic indices. The hemispheric approach may be particularly promising for sunspot-number predictions, where hemispheric observations are directly available and the additional uncertainty associated with reconstructed radio flux series can be avoided. Beyond methodological development, the proposed approach is suitable for implementation in real-time forecasting systems and space-weather operations. Multi-year radio flux predictions enabled by this approach can contribute to assessments of the near-Earth radiation environment by improving the characterization of solar-cycle-driven galactic cosmic ray modulation. They can also support re-entry predictions, orbital-lifetime estimates, and long-term mission planning by providing improved estimates of future thermospheric conditions over timescales relevant to spacecraft operations.

\begin{acknowledgments}
The authors acknowledge the Ottawa and Dominion Radio Astrophysical Observatory (Penticton, Canada) for providing the F10.7 index, the Toyokawa and Nobeyama radio observatories (Japan) for providing the F30 index, and WDC-SILSO at the Royal Observatory of Belgium (ROB) for the sunspot number data sets. We thank the reviewer for the valuable comments on this study.
\end{acknowledgments}

\begin{contribution}
Z.A. and T.P. developed the method and led the writing of the paper. A.M.V., S.J.H. and A.P. contributed to conceptualization, study development, analysis, and writing. All authors discussed the results and provided feedback on the manuscript.
\end{contribution}

\appendix
\section{Noise statistics identification}\label{appendix_NSI}

The Kalman filter requires the variances of two noise components. The model noise describes uncertainties in the predicted state evolution, while the observation noise describes monthly fluctuations relative to the smoothed flux. We estimate these variances from historical radio-flux records using the noise statistics identification method of \citet{Podladchikova2012}. The method separates the two contributions through analytical relations between their variances and the statistics of one- and two-step observation residuals. Here, we develop an updated implementation in which these residuals are formed using transition coefficients derived from historical M\&L predictions. The identified model noise therefore characterizes uncertainties in the M\&L dynamics used by the filter.

In Equation~(\ref{eq:state_space_model}), $w_j$ and $\eta_j$ represent the noise terms in the state and observation equations, respectively. We assume that the variances of noises $w_j$ and $\eta_j$ are proportional to the smoothed flux,
\begin{equation}
\sigma_{w,j}^{2}=a_{w,j}F_j,
\qquad
\sigma_{\eta,j}^{2}=a_{\eta,j}F_j.
\label{eq:NSI_scaling}
\end{equation}
The coefficients $a_{w,j}$ and $a_{\eta,j}$ are identified separately for each assimilation step $j$, with $j=1,\ldots,6$ for 13-month smoothing and $j=1,\ldots,17$ for 35-month smoothing. This allows the noise statistics to vary across the assimilation interval, while the proportionality to $F_j$ accounts for their dependence on the level of solar activity.
The transition coefficient $\Phi_{j,j-1}$ is constructed directly from historical M\&L predictions according to Equation~(\ref{eq:KF_transition}), with $F_0^{ML}=F_0$ for each prediction trajectory. Here, $j$ denotes the number of months from the last available smoothed value.

To identify coefficients $a_{w,j}$ and $a_{\eta,j}$, we construct squared one- and two-step observation residuals,
\begin{equation}
\begin{aligned}
B_{1,j}
&=\left(F_j^{\mathrm{obs}}
-\Phi_{j,j-1}F_{j-1}^{\mathrm{obs}}\right)^2,\\
B_{2,j}
&=\left(F_j^{\mathrm{obs}}
-\Phi_{j,j-2}F_{j-2}^{\mathrm{obs}}\right)^2,
\end{aligned}
\label{eq:NSI_residuals}
\end{equation}
Here, $\Phi_{j,j-2}=\Phi_{j,j-1}\Phi_{j-1,j-2}$, with both transitions taken from the same M\&L trajectory. Under the assumed noise model, the mathematical expectations of these residuals are
\begin{equation}
\begin{aligned}
\mathrm{E}[B_{1,j}]
&=\sigma_{w,j}^{2}+\sigma_{\eta,j}^{2}
+\Phi_{j,j-1}^{2}\sigma_{\eta_{j-1}}^{2},\\
\mathrm{E}[B_{2,j}]
&=\sigma_{w,j}^{2}
+\Phi_{j,j-1}^{2}\sigma_{w_{j-1}}^{2}
+\sigma_{\eta,j}^{2}
+\Phi_{j,j-2}^{2}\sigma_{\eta_{j-2}}^{2}.
\end{aligned}
\label{eq:NSI_moments}
\end{equation}
These mathematical expectations are evaluated using the transition coefficients defined above and are expressed in terms of the unknown model and observation noise variances. The relations distinguish the two noise components through their different contributions to the one- and two-step residuals.
Substituting Equation~(\ref{eq:NSI_scaling}) and estimating the corresponding mathematical expectations over historical prediction origins provides the coefficients listed in Table~\ref{tab:NSI_coefficients}.

\begin{table*}
\centering
\caption{Identified coefficients of the model and observation noise variances in Equation~(\ref{eq:NSI_scaling}), listed by assimilation step $j$. The coefficients have units of sfu when flux is expressed in sfu. Dashes indicate steps not used for 13-month smoothing.}
\label{tab:NSI_coefficients}
\begin{tabular}{c*{8}{c}}
\toprule
& \multicolumn{4}{c}{13-month smoothing}
& \multicolumn{4}{c}{35-month smoothing}\\
\cmidrule(lr){2-5}\cmidrule(lr){6-9}
& \multicolumn{2}{c}{F10.7}
& \multicolumn{2}{c}{F30}
& \multicolumn{2}{c}{F10.7}
& \multicolumn{2}{c}{F30}\\
\cmidrule(lr){2-3}\cmidrule(lr){4-5}
\cmidrule(lr){6-7}\cmidrule(lr){8-9}
$j$ & $a_{w,j}$ & $a_{\eta,j}$
& $a_{w,j}$ & $a_{\eta,j}$
& $a_{w,j}$ & $a_{\eta,j}$
& $a_{w,j}$ & $a_{\eta,j}$\\
\midrule
1  & 0.035 & 0.643 & 0.020 & 0.196 & 0.039 & 0.641 & 0.019 & 0.198\\
2  & 0.036 & 0.642 & 0.020 & 0.196 & 0.040 & 0.640 & 0.019 & 0.198\\
3  & 0.037 & 0.641 & 0.020 & 0.196 & 0.042 & 0.639 & 0.020 & 0.197\\
4  & 0.038 & 0.641 & 0.021 & 0.195 & 0.044 & 0.638 & 0.020 & 0.197\\
5  & 0.040 & 0.640 & 0.021 & 0.195 & 0.046 & 0.638 & 0.022 & 0.196\\
6  & 0.041 & 0.641 & 0.022 & 0.195 & 0.048 & 0.638 & 0.022 & 0.196\\
7  & -- & -- & -- & -- & 0.050 & 0.637 & 0.023 & 0.196\\
8  & -- & -- & -- & -- & 0.051 & 0.638 & 0.025 & 0.196\\
9  & -- & -- & -- & -- & 0.052 & 0.638 & 0.025 & 0.196\\
10 & -- & -- & -- & -- & 0.053 & 0.639 & 0.026 & 0.196\\
11 & -- & -- & -- & -- & 0.053 & 0.639 & 0.027 & 0.197\\
12 & -- & -- & -- & -- & 0.053 & 0.640 & 0.027 & 0.197\\
13 & -- & -- & -- & -- & 0.053 & 0.640 & 0.026 & 0.197\\
14 & -- & -- & -- & -- & 0.051 & 0.641 & 0.026 & 0.197\\
15 & -- & -- & -- & -- & 0.050 & 0.642 & 0.027 & 0.198\\
16 & -- & -- & -- & -- & 0.048 & 0.643 & 0.025 & 0.198\\
17 & -- & -- & -- & -- & 0.049 & 0.643 & 0.024 & 0.198\\
\bottomrule
\end{tabular}
\end{table*}

Reliable estimation requires a valid noise model, identifiable coefficients, and a sufficiently large historical sample for stable estimation of the expected one- and two-step residuals. The use of M\&L transitions makes the identified uncertainties directly relevant to the dynamics employed in the filter. Incorporating these uncertainties into Equations~(\ref{eq:KF_prediction})--(\ref{eq:KF_update}) supports a better-calibrated balance between the initial prediction and monthly observations, providing a basis for more reliable estimates of the current smoothed flux.

\bibliography{My_References}{}
\bibliographystyle{aasjournalv7}

\end{document}